\pdfoutput=1

\documentclass[aps,prx,twocolumn,superscriptaddress,longbibliography]{revtex4-1}

\usepackage{graphicx,bm,amsmath,eufrak,amssymb,url}
\usepackage[usenames,dvipsnames]{color}
\usepackage[colorlinks,bookmarks=false,citecolor=NavyBlue,linkcolor=Red,urlcolor=blue]{hyperref}
\usepackage[bbgreekl]{mathbbol}
\usepackage{xcolor}
\usepackage{hyperref}

\usepackage{subcaption}

\usepackage[T1]{fontenc}
\usepackage[latin9]{inputenc}
\usepackage[english]{babel} 
\usepackage{verbatim}
\usepackage{graphicx}

\usepackage{bbold}
\newcommand{\Id}{{\mathbb 1}}

\newcommand{\be}{\begin{equation}}
\newcommand{\ee}{\end{equation}}
\newcommand{\bea}{\begin{eqnarray}}
\newcommand{\eea}{\end{eqnarray}}

\newcommand{\vac}{|\text{\o}\rangle}
\newcommand{\gr}{|\!\rightarrow\rangle}
\newcommand{\gl}{|\!\leftarrow\rangle}
\newcommand{\gu}{|\!\uparrow\rangle}
\newcommand{\gd}{|\!\downarrow\rangle}

\definecolor{smoothred}{HTML}{C5232F}
\definecolor{mygreen}{rgb}{0,0.5,0}
\definecolor{myblue}{rgb}{0,0,0.75}
\definecolor{mymagenta}{cmyk}{0,1,0,0.12}
\definecolor{myindigo}{HTML}{6B00C2}

\def\doi{http://dx.doi.org/}

\begin{document}

\title{Two-dimensional quantum-link lattice Quantum Electrodynamics at finite density}

\author{Timo Felser}
\affiliation{Theoretische Physik, Universit\"at des Saarlandes, D-66123 Saarbr\"ucken, Germany.}
\affiliation{Dipartimento di Fisica e Astronomia ``G. Galilei'', Universit\`a di Padova, I-35131 Padova, Italy.}
\affiliation{Istituto Nazionale di Fisica Nucleare (INFN), Sezione di Padova, I-35131 Padova, Italy.}   

\author{Pietro Silvi}
\affiliation{Center for Quantum Physics, and Institute for Experimental Physics, 
University of Innsbruck, A-6020 Innsbruck, Austria.}
\affiliation{Institute for Quantum Optics and Quantum Information, 
Austrian Academy of Sciences, A-6020 Innsbruck, Austria.}

\author{Mario Collura}
\affiliation{Theoretische Physik, Universit\"at des Saarlandes, D-66123 Saarbr\"ucken, Germany.}
\affiliation{Dipartimento di Fisica e Astronomia ``G. Galilei'', Universit\`a di Padova, I-35131 Padova, Italy.}
\affiliation{SISSA -- International School for Advanced Studies, I-34136 Trieste, Italy.}

\author{Simone Montangero}
\affiliation{Dipartimento di Fisica e Astronomia ``G. Galilei'', Universit\`a di Padova, I-35131 Padova, Italy.}   
\affiliation{Istituto Nazionale di Fisica Nucleare (INFN), Sezione di Padova, I-35131 Padova, Italy.}   

\date{\today}

\begin{abstract}
We present an unconstrained tree tensor network approach to the study of lattice gauge theories in two spatial dimensions showing how to perform numerical simulations of theories in presence of fermionic matter and four-body magnetic terms, at zero and finite density, with periodic and open boundary conditions.  
We exploit the quantum link representation of the gauge fields and demonstrate that a fermionic rishon representation of the quantum links allows us to efficiently handle the fermionic matter while finite densities are naturally enclosed in the tensor network description.  
We explicit perform calculations for quantum electrodynamics in the spin-one quantum link representation on lattice sizes of up to 16x16 sites, detecting and characterizing different quantum regimes. In particular, at finite density, we detect signatures of a phase separation 
as a function of the bare mass values at different filling densities. 
The presented approach can be extended straightforwardly to three spatial dimensions.

\end{abstract}

\maketitle


Recent progress in quantum simulations are paving the way to the possibility 
of studying high-energy physics phenomena with tools developed in low-energy quantum physics~\cite{zcr12,bdmrswz12,Jordan12,tczl13,bbdrswz13,zcr13,tcoml13,mrsels15,wiese13,zcr16,martinez16,banuls19,banuls19-1}.
In the Standard Model, forces are mediated through gauge fields, 
thus gauge invariant field theories -- e.g., quantum electrodynamics (QED) for the Abelian case
or quantum chromodynamics (QCD) for the non-Abelian scenario -- 
are fundamental building blocks to our understanding 
of all microscopic processes ruling the dynamics of elementary particles~\cite{gauge1,gauge2}. 
When discretizing the gauge theories, the dynamical gauge variables obey a lattice formulation of the original quantum field theory which is referred to as a Lattice Gauge Theory (LGT)~\cite{ks75,frad13}. LGTs encode many-body interactions satisfying exact constraints, encoding a lattice-discretized version of the local gauge invariance, e.g., in QED the Gauss's law $\nabla\cdot E = 4\pi \rho$.  Many of the collective phenomena arising from these theories, including the phase diagram, have yet to be fully characterized \cite{brambilla2014},
especially for higher spatial dimensions at finite charge density.

Possibly the most successful tool to investigate LGTs are Montecarlo simulations based on lattice formulations~\cite{s62,w74,ks75,k79,k83,hwo97}. However, the Montecarlo approach suffers from the infamous sign problem for complex actions, e.g., at finite fermion density (matter/antimatter unbalance), which naturally arises in LGTs~\cite{banuls19-1,tw05}.
Another very promising alternative to simulate lattice gauge theories is based on 
Tensor Network (TN) methods. They have already shown significant capabilities in describing many condensed matter and chemistry problems and to study lattice gauge theories in one spatial dimension~\cite{bsbh02,srcm14,pdrzm16,dm16,bccjs13,bccjk18,pstm19,montangero18,fjk_arxiv,bjk19,banuls19,tcl14,bhvvv14,rpdzm14,hvscv15,srdtm17,bccjk17,sskbdc_arxiv,pstm19}.
So far, very few attempts have been made 
to capture the phase properties (e.g. at zero temperature)
of a lattice analogue of an Abelian gauge theory 
in higher spatial dimensions~\cite{tczl13,p77,bmk77,dqsw79,b-m79,weimer10,zohar2018tnmc,tmd19,celi19,huang19}, 
none of them in the presence of fermionic matter at finite density.

In this work, we fill this gap and develop a computationally tractable Hamiltonian
formulation of low-energy QED in two spatial dimensions. We show that TN states allow 
for an accurate representation of its many-body ground state, 
thus allowing to identify the different regimes, 
and effectively test the response of the system to a finite density of charge.
The study of lattice gauge Hamiltonians at finite chemical potential 
is in general out-of-reach for Montecarlo-based techniques~\cite{tw05,banuls19-1}:
here we show that using an unconstrained Tree Tensor Network (TTN)~\cite{TTN14} 
and the quantum link formalism of lattice gauge theories~\cite{h81,or90,cw97,bdmrswz12,bcw99}, 
we can face this highly non-trivial setup.
The techniques developed in this paper
not only provide the basic ingredients for an efficient calculation
of the phase diagram of simple lattice gauge models,
but they can be extended to more complex theories and higher dimensions. 
 
We demonstrate the effectiveness of the presented approach by focusing on the low-energy properties, 
both at zero  and finite charge density, 
of a two-dimensional lattice quantum link theory with $U(1)$ gauge symmetry.  
Specifically, we investigate a model involving (spinless, flavorless) Kogut-Susskind matter fermions~\cite{ks75,k79} and 
$U(1)$ electromagnetic gauge fields, truncated to a Spin-$S$ compact representation.
Hereafter, we set $S=1$, the smallest representation where all Hamiltonian terms are non-trivial. 
The calculations for higher spin representations are numerically demanding but straightforward.
We investigate the (zero temperature) phase diagram in the zero global charge scenario
without and with finite magnetic coupling.
We observe that, both magnetic and electric Hamiltonian terms, separately, 
hinder the creation of a charge-crystal configuration, 
which emerges at large negative bare masses. However, when electric and magnetic terms are mutually frustrated, the charge-crystal is restored.
Moreover, we study the ground state in the presence of a finite
charge density, which we can directly control in the TN ansatz state.
Small charge densities impact the zero charge phases as follows: in the {\it vacuum regime} charges 
aggregate at the system (open) boundaries, suggesting the existence of a spatial phase separation between the bulk and the boundaries;
this is reminiscent of the classical electrodynamics properties of a perfect conductor, where $\nabla \cdot E = 0$ in the bulk
and the excess of charge is redistributed on the outer surface of the conductor. 
On the contrary, the {\it charge-crystal regime}, which is full of matter/antimatter, is characterized by a homogeneous delocalization of the charge-hole, resulting into a quasi-flat charge distribution in the bulk, therefore reminiscent of a plasma phase~\cite{plasma}.

Finally, we stress that the quantum link formulation provides the ideal tools to establish a connection between LGTs and atomic lattice experiments~\cite{bdz08,hart18}, 
In this framework, the dynamical gauge fields are usually represented by spin degrees of freedom, 
which have a natural mapping to typical condensed-matter models, like Hubbard Hamiltonians or locally constrained Ising-like Hamiltonians. 
These models can be engineered with cold atoms in optical lattices~\cite{tcoml13,martinez16},
or within the very promising experimental setups involving Rydberg atom chains~\cite{s_etal19,ncm19} and can be 
straightforwardly numerically simulated with the presented techniques, to verify and benchmark the experimental results and 
to carefully and quantitatively compare the limits, the precision and the efficiencies of the classical and quantum simulations. 

 The paper is structured as follows: 
 In Sec.~\ref{sec:model}, we present the 2D lattice gauge Hamiltonian and its quantum link formulation in terms of the gauge field Spin-$1$
 compact representation. We also give some technical details of the Tensor Network numerical simulations. 
 In Sec.~\ref{sec:zero_charge} we focus on the ground-state properties in the zero-charge sector:
 we explore the phase space of the model by varying the mass and the electric coupling; 
 we then analyze the effect of a finite magnetic coupling.
 Sec.~\ref{sec:finite_charge} is devoted to study the equilibrium properties at finite 
 charge density. We exploit TTN techniques to investigate how the charges redistribute all over the lattice depending on the Hamiltonian couplings.
 Finally, we draw our conclusions in Sec.~\ref{sec:conclusion}
 and give additional supplementary technical details in the Appendices.
 
\begin{figure}[t!]
\includegraphics[width=\columnwidth]{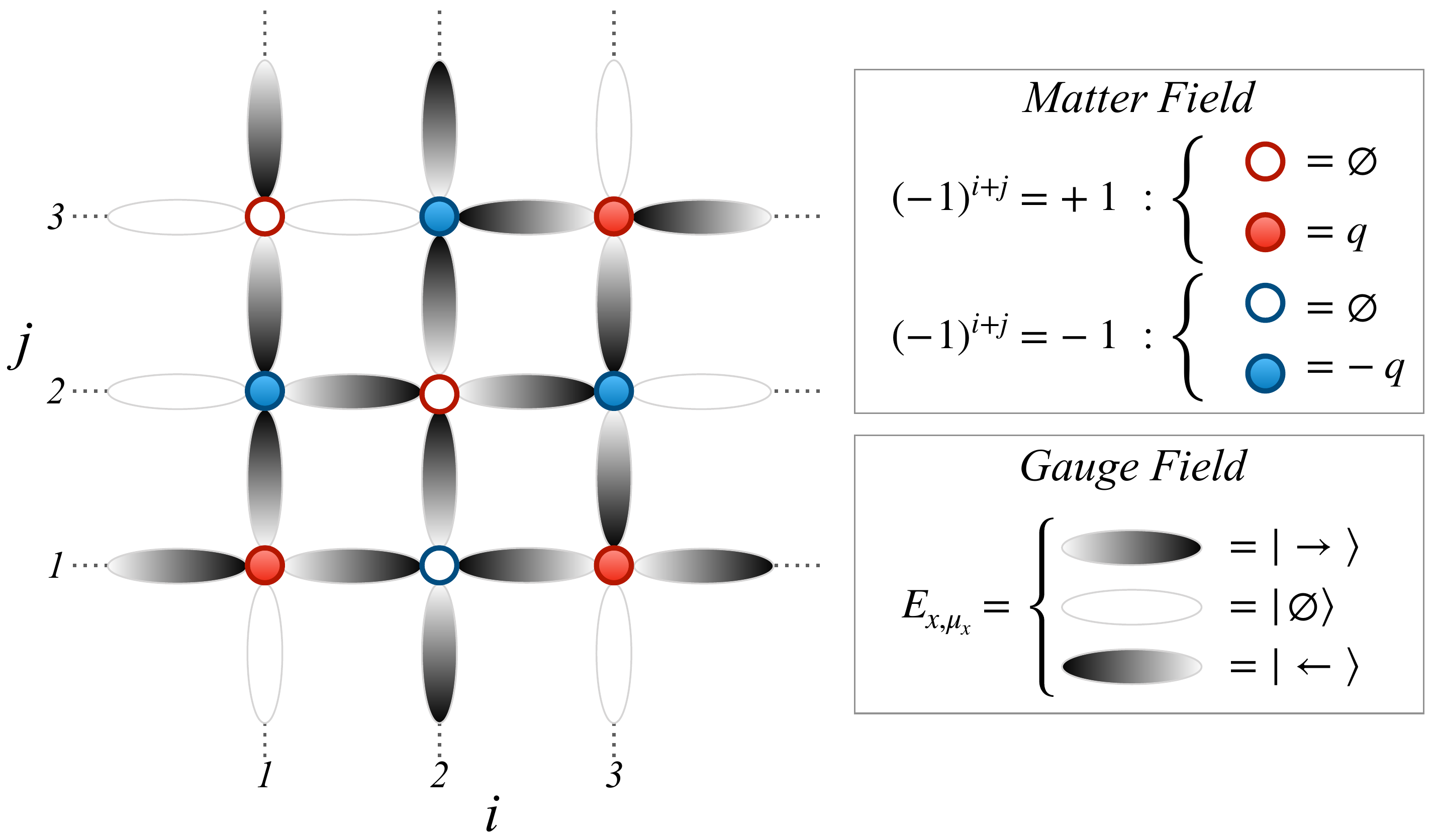}
\caption{ \label{fig:sketch1}
Sketch of the 2D lattice gauge theory in the Spin-$1$
representation. The cartoon of the square lattice shows a specific gauge-invariant
configuration of the matter and gauge fields with zero total charge ($3$ particles and $3$ antiparticles).
Staggered fermions represent matter and antimatter fields on a lattice bipartition: 
on the even (odd) bipartition, a full red (blue) site represents a particle (antiparticle).
The gauge field points in the positive direction of the color gradient. 
}
\end{figure}

\section{Model and Methods}\label{sec:model}
We consider a field theory on a 2D square lattice with $U(1)$ local gauge symmetry.
The sites of a finite $L$$\times$$L$ square lattice host the matter field, while the quantum gauge field lives
on the lattice links, with open boundary conditions.
Following the Kogut-Susskind (staggered)
formulation~\cite{ks75,k79}, the discretization of the matter field is performed by introducing a staggered {fermionic} field, 
whose positive energy solutions lie on the even sites and negative ones on the odd sites. 
The matter field is thus described by spinless, flavorless Dirac fermions, whose operator algebra satisfies 
the usual canonical anti-commutation relations 
$\{\hat\psi_{x},\hat\psi^{\dag}_{x'}\} = \delta_{x,x'}$. 
In particular, in the even sub-lattice
particles represent fermions with electric charge $+q$ (`positrons'), 
whilst in the odd sub-lattice holes represent anti-fermions with electric charge $-q$ (`electrons').
Here a lattice site $x$ labels a 2D coordinate $x\equiv (i,j)$,
and the parity $p_{x}\equiv (-1)^{x} = (-1)^{i+j}$ of a site, distinguishing the two sub-lattices, 
is well defined on the square lattice (see Fig.~\ref{fig:sketch1}).

The gauge field is defined on the lattice links, 
its algebra constructed by the electric field operator 
$\hat E_{x,\mu} = \hat E^{\dagger}_{x,\mu}$ 
and its associated parallel transporter
$\hat U_{x,\mu}$, which is unitary $U_{x,\mu} U^{\dagger}_{x,\mu} = \Id$
and satisfies $[\hat E_{x,\mu},\hat U_{y,\nu}] = \delta_{x,y}\delta_{\mu,\nu}\hat U_{x,\mu}$.
Here $\mu$ ($\nu$) represents the positive unit lattice vector in one of the two orthogonal directions, namely 
$\mu_x \equiv(1,0)$ and $\mu_y \equiv(0,1)$,
thus $(x,\mu)$ uniquely defines a link.
For comfort of notation, we also allow (technically redundant) negative unit lattice vectors $-\mu_x$ and $-\mu_y$,
with the convention $\hat E_{x+\mu,-\mu} = - \hat E_{x,\mu}$, and in turn $\hat U_{x+\mu,-\mu} = \hat U_{x,\mu}^{\dag}$.
With such definitions, apart from a rescaling due to the lattice spacing regularization~\cite{ks75,k79}, 
the two-dimensional lattice \text{QED} Hamiltonian, including a magnetic plaquette term, reads
\bea\label{eq:H_QED}
 \hat H & = & 
 - t \sum_{x,\mu}
  \left( \hat\psi^{\dag}_{x} \, \hat U_{x,\mu} \, \hat\psi_{x+\mu} + \text{H.c.} \right) \nonumber \\
  & + & m \sum_{x}(-1)^{x} \hat\psi^{\dag}_{x}\hat\psi_{x}  +  \frac{g^{2}_{e}}{2} \sum_{ x,\mu} \hat E_{x,\mu}^{2}  \\
 &- & \frac{g^{2}_{m}}{2}\sum_{x}  
 \left(\hat U_{x,\mu_{x}}  \hat U_{x+\mu_{x},\mu_{y}}  
 \hat U^{\dag}_{x+\mu_{y},\mu_{x}}  \hat U^{\dag}_{x,\mu_{y}} + \text{H.c.} \right) \nonumber
\eea
where the coordinate $\mu$ runs in $\{\mu_{x},\mu_{y}\}$.
The first term in Eq. \eqref{eq:H_QED} provides the minimal coupling between gauge and matter fields
associated with the coupling strength $t$.
It describes a process of particle-antiparticle pair creation/annihilation, 
where the parallel transporter operator guarantees that the local gauge symmetries are not violated.
The second term in the Hamiltonian represents the energy associated to the fermionic bare mass, 
and it appears as a staggered chemical potential according to the Kogut-Susskind prescription.
For numerical purpose, it has been redefined by adding 
an overall constant $ m L^2/2$, thus replacing
$
(-1)^{x} \hat\psi^{\dag}_{x}\hat\psi_{x}
\to
\delta_{x,e} \hat\psi^{\dag}_{x}\hat\psi_{x} + \delta_{x,o}\hat\psi_{x}\hat\psi^{\dag}_{x}
$ (see Appendix~\ref{app:matrix}).
This way, a filled local state in the even sub-lattice 
cost positive energy $m$ and carries charge $q$; 
otherwise, when an odd site is empty, the energy cost is still $m$,
but it corresponds to having an antiparticle (a hole) with charge $-q$.
The last two terms contribute to the gauge field dynamics:
the electric part with coupling $g_e$, is completely local. The magnetic part, with coupling $g_m$ instead,
is constructed by considering the smallest Wilson loop 
-- product of parallel transporter $\hat U_{x,\mu}$ in a closed loop -- the size of a plaquette.
Its name is related to the fact that it generates the magnetic contribution to the energy density in the continuum limit.

The LGT Hamiltonian $\hat H$ commutes with 
the local Gauss's law generators (in unit of $q$)
\be\label{eq:gauss}
\hat G_{x} =   \hat\psi^{\dag}_{x}\hat\psi_{x} - \frac{1-p_x}{2} 
- \sum_{\mu} \hat E_{x,\mu},
\ee
where the unit lattice vector $\mu$ in the sum runs in 
$\{\pm \mu_{x},\pm\mu_{y}\}$, while $p_x = (-1)^x$ is, again, the lattice site parity.
In addition, the model exhibits an $U(1)$ global symmetry, 
namely the conservation of the total charge  
$\hat Q = \sum_{x}[\hat\psi^{\dag}_{x}\hat\psi_{x} - \frac{1-p_x}{2} ] = - \frac{L^2}{2} + \hat{N}$, 
equivalent, apart from a constant, to the number conservation 
$\hat{N} = \sum_x \hat\psi^{\dag}_{x}\hat\psi_{x}$ of Kogut-Susskind matter fermions.
As a consequence of the convention, using $\hat E_{x,-\mu} = - \hat E_{x-\mu,\mu}$, 
the sum of all $4$ terms of the gauge field around the lattice site $x$ corresponds to the outgoing electric flux, i.e. 
$
\sum_{\mu} \hat E_{x,\mu} = E_{x,\mu_x}+E_{x,\mu_y} - E_{x-\mu_x,\mu_x}-E_{x-\mu_y,\mu_y}.
$
The gauge invariant Hilbert space is thus given by
all states $|\Phi\rangle$ satisfying $\hat G_{x}|\Phi\rangle = 0$ at every site $x$.
As each electric field degree of freedom is shared by two Gauss' generators $G_x$, 
the generators themselves overlap, and projecting onto the gauge-invariant subspace becomes a non-local operation.
Only for 1D lattice QED, or lattice Schwinger model~\cite{s62}, it is possible to integrate out the gauge variables 
and work with the matter field only (albeit with long-range interactions)~\cite{bccj13}.
However, in two dimensions,
a given (integer occupation) realization of the matter fermions does not fix 
a unique gauge field configuration, thus requiring explicit treatment of the gauge fields as quantum variables.
A numerically-relevant complication, related to
the standard Wilson formulation of lattice gauge theories, 
arises from the gauge field algebra, 
$[\hat{E},\hat{U}] = \hat{U}$ with $\hat{E} = \hat{E}^{\dagger}$ 
and $\hat{U} \hat{U}^{\dagger} = \hat{U}^{\dagger} \hat{U} = \Id$,
whose representations are always infinite dimensional.
Simply put, if a representation contains the gauge field state $|\alpha\rangle$, 
such that $\hat{E}|\alpha\rangle = \alpha |\alpha\rangle$ with $\alpha \in \mathbb{R}$, 
then the states $|\alpha \pm 1\rangle = \hat{U}^{\pm 1} |\alpha\rangle$ belong to the representation as well.
By induction, the representation must contain all the states $|\alpha+\mathbb{N}\rangle$, 
which are mutually orthogonal as distinct eigenstates of $\hat{E}$, thus
the representation space dimension is at least countable infinite.

In order to make the Hamiltonian numerically tractable via Tensor Network methods,
we need to truncate the local gauge field space to a finite dimension. 
For bosonic models, this is typically done by introducing an energy cutoff and eliminating states 
with single-body energy density beyond it, while checking {\it a posteriori} the introduced approximation. 
Similarly, for  U(1) lattice gauge theories, we truncate the electric field according to the quantum link model formulation.
Specifically, the gauge fields are substituted by Spin operators, namely
$
\hat E_{x,\mu} = (\hat S^{z}_{x,\mu} + \alpha)
$
and
$
\hat U_{x,\mu} = \hat S^{+}_{x,\mu} / s,
$
such that $\hat{E}$ is still hermitian and the commutation relation 
$[\hat E_{x,\mu},\hat U_{y,\nu}] = \delta_{x,y}\delta_{\mu,\nu}\hat U_{x,\mu}$
is preserved~\cite{bdmrswz12}, however $\hat{U}$ is no longer unitary
for any {\it finite} spin-$s$ representation $|\hat{\boldsymbol{S}}|^2 = s(s+1) \Id$.
The original algebra is then restored in the large spin limit $s \to \infty$, 
for any {\it background} field $\alpha \in \mathbb{R}$.
Similar truncation strategies, based on group representations, 
can be applied to non-Abelian gauge theories as well~\cite{tcl14,zb15}.
In the following, we make use of the Spin-$1$ representation ($s=1$),
under zero background field $\alpha = 0$,
which captures reasonably well the low-energy physics 
of the theory, especially in the parameter regions wherein the ground-state 
is characterized by small fluctuations above the bare vacuum.
$s=1$ is the smallest spin representation exhibiting a nontrivial  
electric energy contribution. 
In fact, for $s=1/2$, we have that $\hat{E}_{x,\mu}^2 \propto (\sigma^{z}_{x,\mu})^2 = \Id$ 
is simply a constant in the Hamiltonian, thus $g_e^2$ plays no role.
In 1D it was observed that truncated gauge representations converge rapidly to the continuum theory,
e.g. in the Schwinger model~\cite{rpdzm14,notar15,ercol18}, 
reinforcing quantitative validity of the results obtained in the simplified model.
Deviations between the truncated and the full-fledged lattice theory
are expected to arise when $g_m^2$ is the dominant coupling, as we show for a $2 \times 2$
example in Appendix \ref{app:2x2}.

Let us mention that, in the formulation of the lattice QED implemented on our numerical algorithms,
Eq.~\eqref{eq:H_QED}, we consider the respective couplings of the various Hamiltonian terms, namely
$t$, $m$, $g_e^2$ and $g_m^2$ as independent, dimensionless parameters. This is a practical advantage
of the numerical interface allowing us to treat the Hamiltonian terms on equal footage, and, in what
follows, we will set the energy-scale via $t = 1$.
We stress, however, that in the original Hamiltonian formulation of lattice QED \cite{ks75,k79}, these couplings
are mutually related as
$t = \frac{1}{a}$,
$m = m_0$,
$g^{2}_{e} = \frac{g^{2}}{a}$,
$g^{2}_{m} = \frac{8}{g^2 a}$,
where $g$ is the coupling constant of QED, $m_0$ is the matter field bare mass, and $a$ is the lattice spacing
of the lattice discretization. In this sense, {\it physical} realizations of the lattice QED, only depend on two actual
parameters: $m' = m_0 a > 0 $ and $g^{2} > 0 $. Nevertheless, in this work we aim to highlight that our numerical simulations
are not limited to these physical scenarios, and we will keep our effective couplings independent and not bound to positive values.
We leave more detailed convergence analysis along the physical regimes to future work, along the lines of similar studies
already presented for 1D systems~\cite{banuls19-1}.


\subsection{Spin-$1$ compact representation of $U(1)$}\label{sec:S1rep}
In the Spin-$1$ representation, the electric field operator 
allows three orthogonal states for the electric flux 
(in unit of the charge $q$), graphically represented in Fig.~\ref{fig:sketch1}.
For a horizontal link $(x,\mu_x)$ we write the eigenbasis of $E_{x,\mu_{x}}$ as
\[
\hat E_{x,\mu_{x}}\gr = + \gr, ~ \\
\hat E_{x,\mu_{x}}\vac= 0, ~ \\
\hat E_{x,\mu_{x}}\gl = -\gl,
\]
on which the parallel transporter acts as
${\hat U_{x,\mu_{x}}\gr = 0}$, ${\hat U_{x,\mu_{x}}\vac = \gr}$ 
and ${\hat U_{x,\mu_{x}}\gl= \vac}$;
and analogously ${\hat E_{x+\mu_x,-\mu_{x}}\gr= - \gr}$. 
A similar set of states 
can be defined in the vertical links $(x,\mu_y)$,
such that ${\hat E_{x,\mu_{y}}\gu = \gu}$, 
${\hat E_{x,\mu_{y}}\vac= 0}$ 
and ${\hat E_{x,\mu_{y}}\gd = -\gd}$.

In this work we introduce an algebraic technique, similar to the {\it rishon} 
representations common in quantum link models~\cite{h81,or90,cw97,bcw99,bdmrswz12},
which has the advantage of automatically accounting for the Gauss' law, 
while carefully reproducing the anticommutation relations
of the matter fermions without resorting to Jordan-Wigner string terms (see next section). 
This strategy relies on splitting the gauge field space on each link $(x,\mu)$
into a {\it pair} of $3$-hardcore fermionic modes, defined later on. 
We say that each mode in this pair `belongs' to either of the sites sharing the link, in this case $x$ and $x+\mu$.

\begin{figure}[t!]
\includegraphics[width=0.47\textwidth]{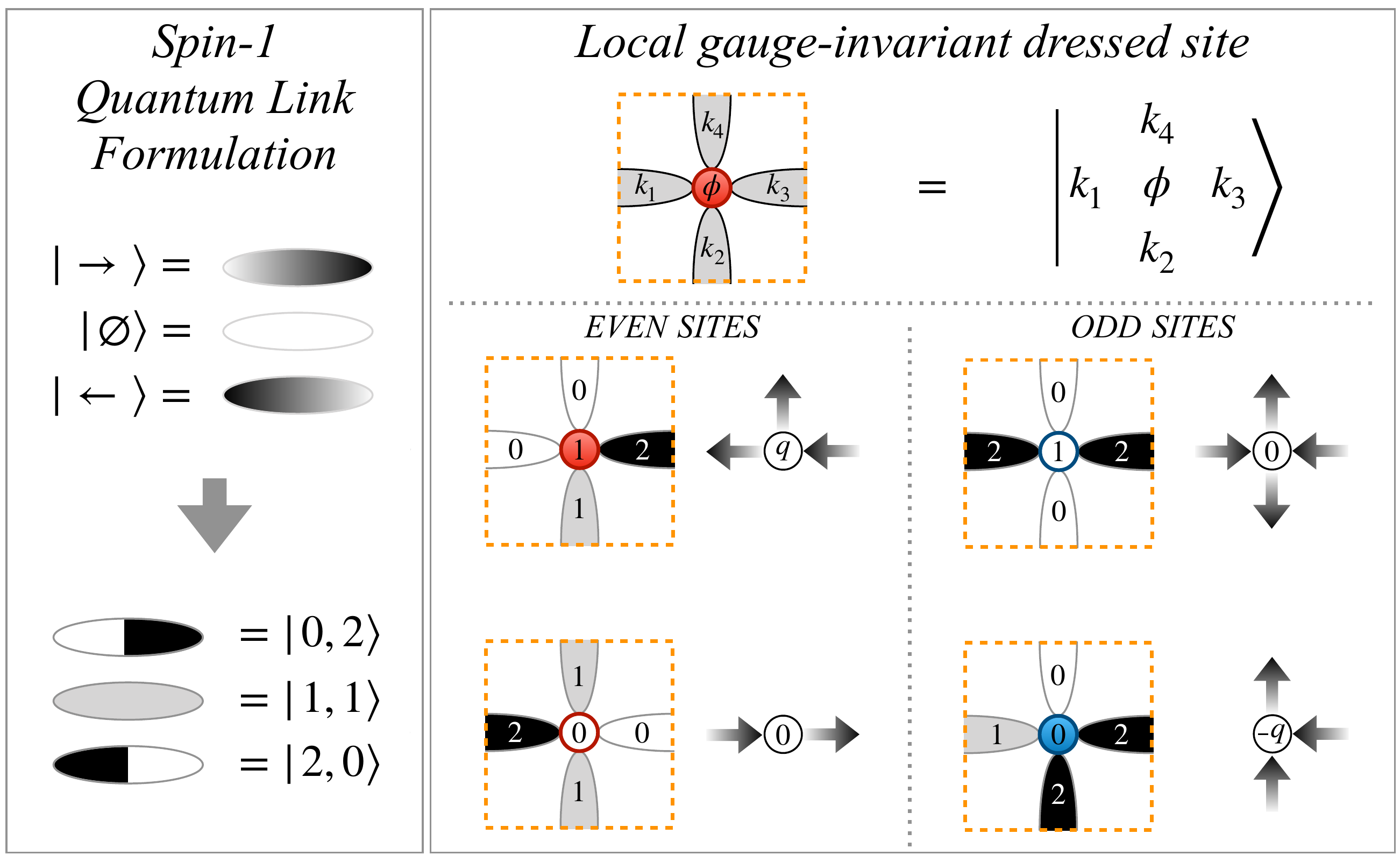}
\caption{ \label{fig:sketch2}
Mapping of the gauge field states in the Spin-$1$
representation to the fermionic Fock states. Each half link is constructed
by employing two species of Dirac fermions. The link symmetry formally
reduces the total number of states to only the $3$ allowed states with $2$ fermions.
We therefore construct the local gauge-invariant dressed site 
by gluing each single matter site together with its neighboring half links.
In the four examples, notice how the quantum number $\phi=1$ represents 
the presence (absence) of a charge (anti-charge) in the even (odd) sites.
}
\end{figure}

Thus, we write
$
{\hat U_{x,\mu} = \hat \eta_{x,\mu}  \hat \eta^{\dag}_{x+\mu,-\mu} }
$,
where the 3-hardcore fermionic operators $\hat{\eta}$ satisfy
$\hat \eta_{x,\mu}^{3} = 0$ (while $\hat \eta_{x,\mu}^{2} \neq 0$) and anticommute
at different positions $\{\hat\eta_{x,\mu}, \hat\eta^{(\dag)}_{y,\nu}\} = 0$,
for $x\neq y$ or $\mu \neq \nu$.
Moreover, these new modes obey anti-commutation relations with the matter field as well, 
i.e. $\{\hat\eta^{(\dag)}_{x,\mu},\hat\psi^{(\dag)}_{y}\}=0$.
To explicitly build this $3$-hardcore fermionic mode $\hat \eta_{x,\mu}$, 
we use two sub-species of Dirac fermions $\hat a_{x,\mu}$
and $\hat b_{x,\mu}$, such that
\be
\hat \eta^{\dag}_{x,\mu} = \hat n^{a}_{x,\mu} \hat b^{\dag}_{x,\mu} 
+ (1- \hat n^{b}_{x,\mu})\hat a^{\dag}_{x,\mu} , \; 
(\hat \eta^{\dag}_{x,\mu})^{2} =  \hat b^{\dag}_{x,\mu} \hat a^{\dag}_{x,\mu},
\label{eq:etadef}
\ee
 where $\hat n^{a}_{x,\mu} = \hat a^{\dag}_{x,\mu} \hat a_{x,\mu}$
 and $\hat n^{b}_{x,\mu} = \hat b^{\dag}_{x,\mu} \hat b_{x,\mu}$
 are the occupation number operators for each sub-species.
This construction provides the local algebra
 \be
[\hat \eta_{x,\mu} , \hat \eta^{\dag}_{x,\mu} ]
=
1-\hat n^{a}_{x,\mu} - \hat n^{b}_{x,\mu},
 \ee
and grants only access to the 3-dimensional subspaces for each 3-hardcore fermion mode,
spanned by the following three states
\be\label{eq:half-link-state}
|0\rangle_{x,\mu}, ~ \;
|1\rangle_{x,\mu} =   \hat a^{\dag}_{x,\mu} |0\rangle_{x,\mu}, ~ \;
|2\rangle_{x,\mu} =   \hat b^{\dag}_{x,\mu} \hat a^{\dag}_{x,\mu} |0\rangle_{x,\mu},
\ee
where $|0\rangle_{x,\mu}$ is the Dirac vacuum of both sub-species,
i.e. $a_{x,\mu} |0\rangle_{x,\mu} = b_{x,\mu} |0\rangle_{x,\mu} = 0$.
The state $\hat b^{\dag}_{x,\mu}  |0\rangle_{x,\mu}$  is disconnected from the other 
three and thus projected away.
Such ``half-link''  local subspace is joined with a similar construction $ \hat a^{\dag}_{x+\mu,-\mu}$
and  $ \hat b^{\dag}_{x+\mu,-\mu}$ on the other half of the link, thus the {\it pair} defines the link space,
and $\hat{E}_{x,\mu}$ will be diagonal in the occupation basis.
While, in principle, a full link space is 3$\times$3$=$9-dimensional,
we can now exploit the following symmetry
\be
\hat L_{x,\mu} = \hat n^{a}_{x,\mu}  + \hat n^{b}_{x,\mu} + \hat n^{a}_{x+\mu,-\mu} + \hat n^{b}_{x+\mu,-\mu},
\ee
which counts the total number of fermions in each link and is a conserved quantity since
$[\hat L_{x,\mu}, \hat E_{x,\mu}] = [\hat L_{x,\mu} ,\hat U_{x,\mu} ] = 0$.
By working in the subspace with two fermions per link, $\hat{L}_{x,\mu} = 2$, 
we reduce the link space to dimension 3 and we can restore the desired algebra. 
First of all, we write the occupation basis (see also Fig.~\ref{fig:sketch2})
\bea\label{eq:link-state}
\gr& = & -|0, 2\rangle = \hat a^{\dag}_{x+\mu,-\mu}  \hat b^{\dag}_{x+\mu,-\mu}  
|0\rangle_{x,\mu} |0\rangle_{x+\mu,-\mu} ,\nonumber \\
\vac & = & |1, 1\rangle =  \hat a^{\dag}_{x,\mu} \hat a^{\dag}_{x+\mu,-\mu}  
|0\rangle_{x,\mu} |0\rangle_{x+\mu,-\mu} ,\\
\gl & = & |2, 0\rangle =  \hat b^{\dag}_{x,\mu}  \hat a^{\dag}_{x,\mu}  
|0\rangle_{x,\mu} |0\rangle_{x+\mu,-\mu}, \nonumber
\eea
so that $\hat{U}_{x,\mu}$ acts correctly, i.e.~where the minus sign in the first equation ensures
${\hat U_{x,\mu}\vac = \gr}$ and ${\hat U^{\dag}_{x,\mu}\gr = \vac}$.
Then, we express the electric field operator as the unbalance of
fermions between the two halves of a link, precisely
\be\label{eq:E-link}
\hat E_{x,\mu} = \frac{1}{2}
\left(
 \hat n^{a}_{x+\mu,-\mu} + \hat n^{b}_{x+\mu,-\mu}
 -\hat n^{a}_{x,\mu}  - \hat n^{b}_{x,\mu}
\right),
\ee
implementing the correct action of $E_{x,\mu}$.
It is worth to mention that this formulation can be extended to higher Spin-$j$ representations, 
where each link becomes a pair of $(2j+1)$-hardcore fermions.

\subsection{The local gauge-invariant dressed sites}\label{sec:dressed}
One of the common issues while working with 
a lattice gauge theory, even in the compact representation of the electric field,
is to properly identify the gauge-invariant Hilbert space.
Due to the overlapping of the Gauss' law generators $\hat G_{x}$,
the identification of the correct local basis is highly non-trivial,
especially for dimensions higher than one~\cite{srcm14}.

Using the 3-hardcore fermion pairs language
gives us a shortcut to this issue.
In fact, we are able to recast the Gauss' law generators as non-overlapping operators,
at the price of enforcing the link constraint
${(\hat L_{x,\mu}-2) |\Phi\rangle  = 0}$.
Using this constraint, we can rewrite the electric field operator
in Eq. (\ref{eq:E-link}) taking only the fermionic operators into account, which act
on the ``half-link'' connected to $x$, i.e.
\be\label{eq:E-half-link}
\hat E_{x,\mu} = 1 - \hat n^{a}_{x,\mu}  - \hat n^{b}_{x,\mu} = [\hat\eta_{x,\mu},\hat\eta^{\dag}_{x,\mu}],
\ee
valid in the link-symmetry invariant space ${(\hat L_{x,\mu}-2) |\Phi\rangle  = 0}$. 
As a consequence, in the Hilbert space with 2 Dirac fermions per link,
the Gauss's law generators become strictly local, 
i.e. containing quantum variables belonging solely to site $x$, 
and read 
\be \label{eq:strictlocalgauss}
 \hat G_{x}  =
\hat\psi^{\dag}_{x}\hat\psi_{x} - \frac{1-p_x}{2} - \sum_{\mu} (1-\hat n^{a}_{x,\mu}-\hat n^{b}_{x,\mu}).
\ee
Within this picture, it is easy to identify a local gauge-invariant basis for the {\it dressed} site
\bea\label{eq:dressed-state}
\Bigg|
\begin{matrix}
& k_4 &\\
k_1 & \phi & k_3 \\
& k_2 &\\
\end{matrix}
\Bigg\rangle_{\!\! x} 
& = & (-1)^{\delta_{k_1,2}+\delta_{k_2,2}} \\
&\times &
|\phi\rangle_{x}
|k_{1}\rangle_{x,-\mu_{x}}
|k_{2}\rangle_{x,-\mu_{y}}
|k_{3}\rangle_{x,\mu_{x}}
|k_{4}\rangle_{x,\mu_{y}}, \nonumber
\eea
where $|\phi\rangle_{x} = (\hat\psi^{\dag}_{x})^{\phi }|0\rangle$
for $\phi\in\{0,1\}$ describes the matter content, while $k_{j}\in\{0,1,2\}$ 
selects a state, from those in Eq. (\ref{eq:half-link-state}), for each respective half-link.
The factor $(-1)^{\delta_{k_1,2}+\delta_{k_2,2}}$ accommodates the sign in Eq. (\ref{eq:link-state}).
In this language, the Gauss' law, cast as Eq.~\eqref{eq:strictlocalgauss}, simplifies to
\be\label{eq:gauss-dressed}
 \phi + \sum_{j=1}^{4} k_{j} = 4 +\frac{1-p_x}{2},
\ee
which fixes the total number of fermions in each dressed site, 
specifically 4 in the even sites, and 5 in the odd sites.
Eq. (\ref{eq:gauss-dressed}) actually reduces the Hilbert space dimension of each dressed site from $162$ to $35$, 
and we use these $35$ states as computational basis for tensor network algorithms.

A fundamental feature of this language is that, since the total number of fermions at each dressed site is conserved,
their parity is conserved as well, thus the gauge-invariant model will exhibit no Jordan-Wigner strings (outside
the dressed sites) in the computational basis.
An operative way to show this property is to consider that the Hamiltonian term
$\hat \psi^{\dag}_{x}\hat U_{x,\mu}\hat\psi_{x+\mu}$, decomposes as the product
of $\hat \psi^{\dag}_{x}\hat\eta_{x,\mu}$ and
$\hat\eta^{\dag}_{x+\mu,-\mu} \hat \psi_{x+\mu}$: each of these two factors is local (acts on a single dressed site)
and commutes with the algebra of other dressed sites. The same applies to the magnetic plaquette term.
In conclusion, by working on the dressed-site computational basis, we can employ standard (spin-model-like)
tensor network techniques, without the requirement of keeping track 
of fermionic parity at each site~\cite{cevv10,cobv10,ksvc10,pv10,shzz-arxiv,bpe09, bpe10}. Notice that this construction can be exploited also to perform quantum computations of two-dimensional LGTs.

\subsection{Tensor Network for 2D lattice gauge simulations}\label{sec:TNsim}

In order to numerically simulate the quantum system, 
we use a two-dimensional Tree Tensor Network (TTN) state to represent the many-body wave function~\cite{TTN14, gqh17, TTNvsNN}.
We work in the computational 35-dimensional local basis for each dressed site, 
defined in the previous section, which automatically encodes the Gauss' law.
Operators appearing in the Hamiltonian (\ref{eq:H_QED}) can be cast in this basis,
either as local operators, or acting on a pair or plaquette of neighboring dressed site
(see Appendix~\ref{app:matrix} for the explicit construction).
The extra link symmetry $\hat L_{x,\mu} = 2$ must be enforced at every pair of neighboring sites.
We do so by introducing an energy penalty for all states violating the link constraints. 
This penalty term is included in the optimization by a driven penalty method - similar to an augmented Lagrangian method - 
which is described in more detail in Appendix~\ref{app:TTN}.
Under all other aspects, the TTN algorithm employed here for finding the many-body ground state
follows the prescriptions of Ref.~\cite{TTNA19}.

In the numerical simulations we fixed the energy scale by setting the coupling strength $t=-1$. 
Furthermore, we worked within a sector with a fixed total charge $\hat Q$, 
by using standard techniques for global symmetry conservation in TNs~\cite{TTNA19,m07,sv13}.
We thus characterized the ground-state properties as function of the mass $m$,
the electric coupling $g_{e}$ and the magnetic coupling $g_m$.

In order to exploit the best performances of our
TTN algorithm, we ran simulations on square lattices $L \times L$ 
with the linear length $L$ being a binary power; in particular, 
we considered $L =4,8$ and $16$,
and varied the TN auxiliary dimension (or bond dimension) up to $\chi \sim 300$. Depending on $L$ and the physical parameters, we obtain a convergence precision between $\sim 10^{-2}-10^{-5}$,
sufficient to characterize the ground-state properties.

\begin{figure}[t!]
\includegraphics[width=0.49\textwidth]{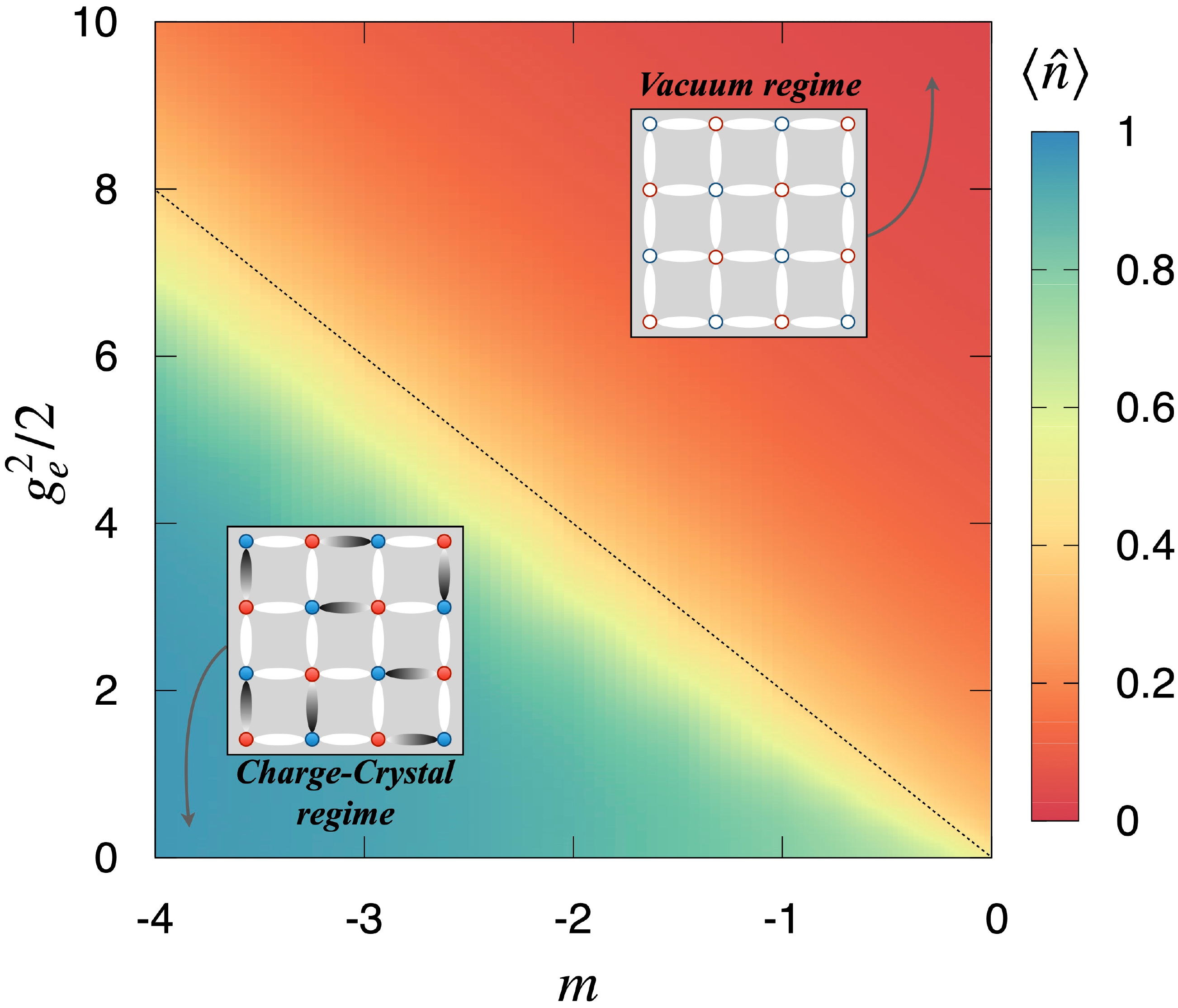}
\caption{ \label{fig:phase_Q0}
Color density plot for $m<0$ obtained from the evaluation of 
the density of matter in the TTN ground state for a $8 \times 8$ lattice system 
with periodic boundary conditions.
The insets are schematic representations of the ground state deep in the two regimes:
the bare vacuum for $g^2_e/2 \gg 2|m|$, 
a typical dimer configuration for $g^2_e/2 \ll -2m$.
The dashed line is located at the classical ($t=0$) transition $g^2_e/2 = -2m$. 
}
\end{figure}

\begin{figure*}[t!]
\includegraphics[width=0.95\textwidth]{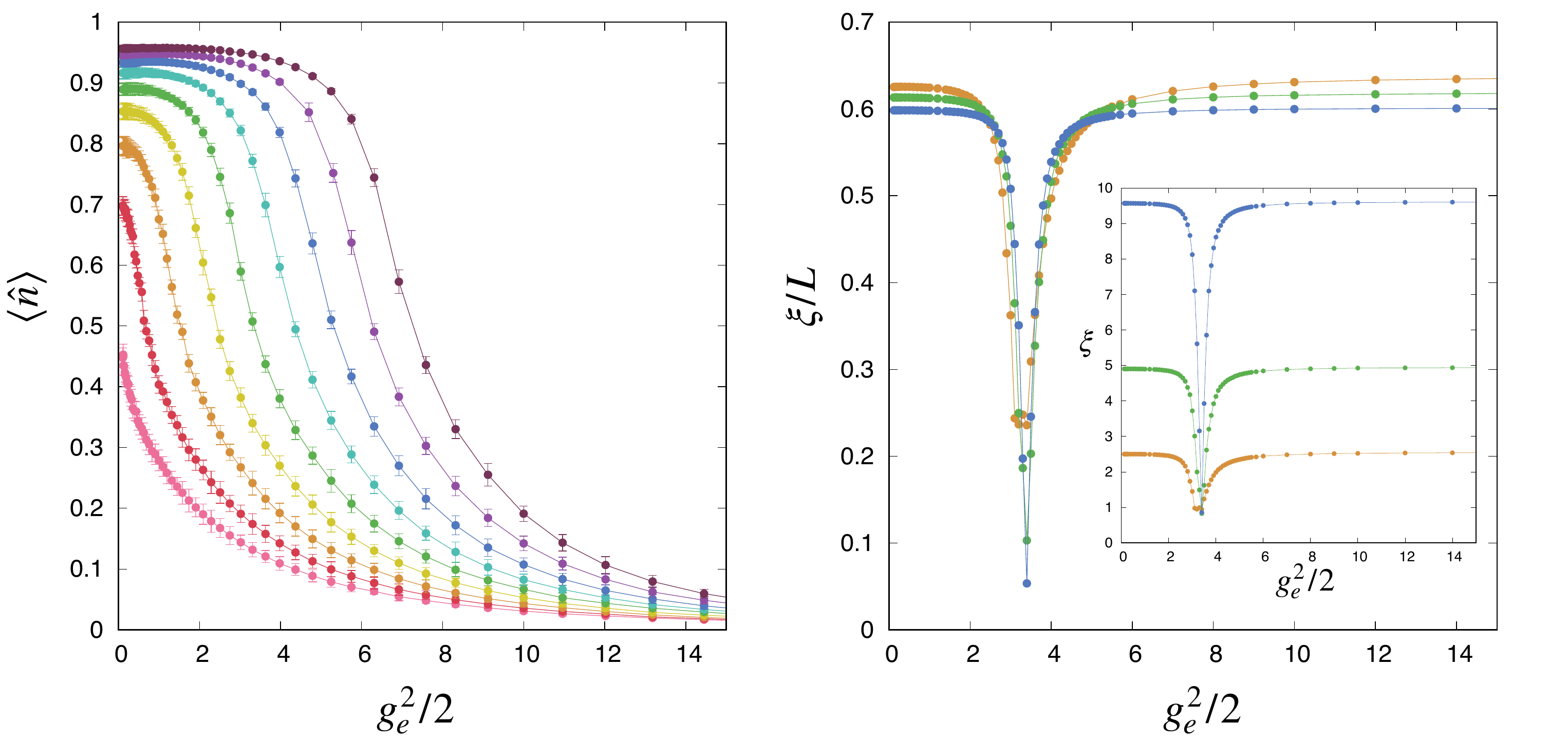}
\caption{\label{fig:Q0_density}\label{fig:Q0_corr}
({\bf left}) Profiles of the matter density as function of the electric field coupling
for different values of negative mass obtained by vertically cutting the color plot
in Fig.~\ref{fig:phase_Q0}. From bottom (pink circles) to top (purple circles) 
mass is taking the values $ m \in \{-0.01, \, -0.5, \, -1, \, -1.5, \, -2, \, -2.5, \, -3, \, -3.5, \, -4\}$. 
({\bf right}) The correlation length, in the ground-state for $m=-2$ and varying the electric coupling,
extracted form the density-density correlation function as explained in the main text. 
Different colors represent different sizes: $4\times 4$ (orange), $8 \times 8$ (green) and $16 \times 16$ (blue). 
}
\end{figure*}

\section{Zero charge density sector}\label{sec:zero_charge}
In this section, we focus on the zero charge density sector $\rho \equiv  \langle \hat Q\rangle/L^2 = 0$,
where there is a balance between matter and antimatter,
and analyze the ground state of Hamiltonian (\ref{eq:H_QED}) within this subspace.
Unless otherwise stated, we consider periodic boundary conditions.
We characterize the ground state of the Hamiltonian
by looking at the energy density $\langle \hat H \rangle/L^2$,
and the particle density 
$\langle \hat n \rangle = \frac{1}{L^2} \sum_{x} \langle \hat{n}_x \rangle$ 
where 
$\hat{n}_x = ( \delta_{x,e} \hat\psi^{\dag}_{x}\hat\psi_{x}
+ \delta_{x,o}\hat\psi_{x}\hat\psi^{\dag}_{x} )$
counts how many charges are in the system, both positive and negative, 
i.e. fermions in even sites plus holes in odd sites.
We start our analysis by first  focusing on the case in which 
the magnetic coupling has been set to zero, $g_m =0$.
Before detailing the numerical results, some analytically-solvable limit cases should be considered.
For large positive values of the bare mass $m \gg t$,
the fluctuations above the bare vacuum are highly suppressed; 
the system exhibits a unique behavior since there is no competition between
the matter term and the electric field term in the Hamiltonian. 
Indeed, to construct pairs of particle/antiparticle, the matter energy  
and the electric field energy both contribute to an overall increasing of the ground-state energy.
In order to explore more interesting phenomena, 
we allow the mass coupling to reach negative values. 
Doing so, we can identify two different regions
depending on the competition between the electric coupling $g^{2}_{e}/2$
and the values of the mass $m < 0$:

{\it (i)} for $g^{2}_{e}/2 \gg 2|m|$, we still have a vacuum-like behavior, 
where we expect a unique non-degenerate ground-state with small particle-density fluctuations.
This regime exists, no matter the value of the mass, as far as the energy cost
to turn on a non-vanishing electric field on a single link 
overcomes the gain in creating the associated pairs of particle/antiparticle. 
Indeed, for any value of the mass and $g^{2}_{e}/2 \to \infty$, 
or for $g^{2}_{e}/2 \neq 0$ and $m\to\infty$,
the presence of a finite electric field, or finite particle density, 
is strictly forbidden and the ground-state flows toward the only admissible configuration, 
namely the {\it bare vacuum}.

{\it (ii)} for $-2m \gg  g^{2}_{e}/2 > 0$ the system is characterized by slightly deformed 
particle-antiparticle dimers; this regime of course only exists for negative value of the 
mass and represents the region wherein the energy gain for creating a couple of particle/antiparticle
largely overcomes the associated electric field energy cost. Here the ground-state 
remains highly degenerate as far as the kinetic energy coupling $|t|$ is much smaller
than all the others energy scales (degeneracy being lifted only at the fourth order in $t$).
In particular, for $g^{2}_{e}/2 \neq 0$ and $m\to-\infty$ the ground state
reduces to a completely filled state. In order to minimize the electric field energy,
particles and antiparticles are arranged in $L^2/2$ 
pairs (where we are assuming $L$ even) sharing a single electric flux in between.
All these configurations are energetically equivalent and their degeneracy
corresponds to the number of ways in which a finite quadratic lattice 
(with open or periodic boundary conditions)  can be fully covered 
with given numbers of ``horizontal'' and  ``vertical'' dimers. 
This number scales exponentially with the system size
as $\exp(L^2 C/\pi)$ for $L\to\infty$, with $C\simeq 0.915966$ the Catalan's constant~\cite{dimer}.
For sake of clarity, we stress that such `dimers' are not entangled clusters of matter and gauge fields; 
they are roughly product states.

Let us mention that the case $g_{e} = 0$ with $m\to \infty$ ($m\to -\infty$) is more pathological
since any gauge-field configuration compatible with the vacuum (dimerized) state is admissible, 
provided the Gauss's law is fulfilled. 
In practice, we may draw a generic closed loop with finite electric flux on top of the vacuum state
without modifying its energy; similar gauge loops may be realized on top of the dimerised state, 
provided it is compatible with the occupied links, without changing its energy as well.
All these configurations are gauge-invariant by construction, 
and increase the degeneracy of the ground-state energy sector.

Our numerical results confirm and extend this picture, 
as it can easily be seen in the phase diagram displayed in Fig.~\ref{fig:phase_Q0},
obtained from TTN simulations in a $8 \times 8$ system.
The matter density is roughly zero in the {\it vacuum regime}; 
otherwise, it takes on a finite value whenever the system exhibits ``dimerisation'', 
i.e. in the {\it charge-crystal regime}.
We checked that the numerical data, both the ground-state energy density and
the particle density, show an asymptotic 
tendency toward the perturbative estimates.
Interestingly, the particle density experiences an abrupt change 
mainly in a narrowed region around $m \simeq - g^{2}_{e}/4$,
where the local slope is becoming steeper as the electric coupling 
(and the mass) is approaching zero (see left panel in Fig.~\ref{fig:Q0_density}),
as roughly predicted by perturbation theory and supported by the exact results in the $2 \times 2$ case 
(see Appendices \ref{app:perturbation} and \ref{app:2x2}).

The two regimes exhibit opposite long-range ordering, identified by the order parameter
${\hat{\mathcal{O}}}_{x} = (-1)^{x} (2 \psi^{\dagger}_{x} \psi_{x} - 1)$, which mutually frustrate
in proximity of the transition. As a confirmation of this property, we can track, for instance,
the {\it full} density-density correlation function
$\mathcal{C}_{x,x'} = \langle {\hat{\mathcal{O}}}_{x} {\hat{\mathcal{O}}}_{x'} \rangle$
(as compared to its connected component
$\mathcal{C}^{0}_{x,x'} = \mathcal{C}_{x,x'} - \langle {\hat{\mathcal{O}}}_{x} \rangle \langle {\hat{\mathcal{O}}}_{x'} \rangle$
). We expect both regimes to exhibit an extensive (linear with $L$) full correlation length, while a sudden drop
in such a quantity identifies frustration, and helps us locate the transition with high precision.
Such behavior is shown in the right panel of Fig.~\ref{fig:Q0_density}, where we quantify the full correlation length
via the estimator ${\xi}_{\text{est}}^2 = \sum_{v} D^2(v) \bar{\mathcal{C}}(v) / \sum_{v} \bar{\mathcal{C}}(v)$,
which uses the spatially-averaged correlation function $\bar{\mathcal{C}}(v) = L^{-2} \sum_{x} \mathcal{C}_{x,x+v}$
and the euclidean metric $D^2(v) = v_x^2 + v_y^2$. Such estimator effectively calculates the two-dimensional variance
of $\bar{\mathcal{C}}(v)$ meant as a distribution (further discussed in Appendix~\ref{app:corlenestimator}).
By testing sizes up to $16\times 16$ we observe that
the actual transition point is slightly below
the classical (i.e., $t=0$) position $g^{2}_{e}/2 = -2m$.
Such an outcome confirms out predictions that particle/antiparticle 
fluctuations, induced by a finite value of the hopping amplitude $t$,  
naturally discourage the charge-crystal order. 

This effect emerges already at the second order 
in perturbation theory treatment (see Appendix~\ref{app:perturbation}),
where the crossing point of the two different ground-state energies, 
$E_{v}$ (vacuum) and $E_{d}$ (dimer), 
slightly shifts toward the dimerised configuration.


A relevant physical question is whether
the system undergoes an actual quantum phase transition across the two regions.
Exactly at $t=0$,
when $m$ crosses the critical value $-g^{2}_{e}/{4}$,
the ground-state exhibits an exact level-crossing, 
passing from the bare vacuum to the charge-crystal energy sector. 
In this limit case, the system experiences a trivial {\it first-order phase transition},
since the gauge-field energy term and the matter-field mass term commute between each other.
However, if we tune the mass at the classical critical value $m = -g^{2}_{e}/{4}$, 
a small hopping amplitude $t\neq 0$ is already sufficient to remove such degeneracy: 
namely, the bare-vacuum energy and the charge-crystal energy get modified 
in a different way so that a gap opens between the two sectors. 
At the critical value of the mass, creation/annihilation of 
pairs of particle/antiparticle has no energetic cost, and the ground state energy sector 
is characterized by all possible states with any number of dimers; 
however, creating a pair in the vicinity of the bare vacuum
is more favorable than annihilating a pair on top a fully dimerised state;
at least, this is true for any finite $L$ (see Appendix \ref{app:perturbation} for details).

A crucial insight comes from the features 
of the overlap between the exact ground-state $|GS\rangle$
and the unique bare vacuum $|\Omega\rangle$ 
(see ~Appendix~\ref{app:2x2} for exact results in the $2\times 2$ case).
Indeed, for the $t=0$ trivial case, it experiences a discontinuous transition when passing
from the vacuum sector to the full dimerised sector, suddenly jumping from one to zero.
Interestingly, for fixed system size $L$, we may evaluate such overlap 
in the approximate ground-state $|GS^{(k)}\rangle$ at a given order $k$ in perturbation theory.
The resulting perturbative expansion of the square of the overlap 
$|\langle\Omega| GS^{(k)}\rangle|^2$ changes continuously in the vacuum regime, 
while it remains identically vanishing, 
i.e. $|\langle\Omega| GS^{(k)}\rangle|^2 = 0$, 
when correcting the fully dimerised state up to $k < L^2/2$.
We thus expect the exact overlap to be continuous at the transition point
and identically vanishing in the thermodynamic limit $L\to\infty$.
As a consequence, for any finite $t$, 
no first-order phase-transition occurs, 
and we may have a {\it second-order phase transition}.
Let us stress that, although the perturbative expansion 
of the fully dimerised state does not produce any change in the overlap
with the bare vacuum for $k < L^2/2$,
local observables do experience perturbative modifications,
simply because the state by itself gets modified.
In particular, as a consequence of the Hellmann-Feynman theorem,
the particle density as a function of the mass coupling $m$ coincide with
the derivative of the ground-state energy density, 
$\langle  GS(m) |  \hat n  | GS(m) \rangle  = \partial_m E_{GS}(m)/L^2$.
A second-order phase transition will thus imply continuous 
profiles of the particle density, with a discontinuous or diverging derivative 
at the transition point. In fact, we have numerical evidence that
the matter density changes continuously when going from one phase to the other (see Fig.~\ref{fig:Q0_density}),
however, it remains very hard to infer about its derivative at the transition point.

\begin{figure}[t!]
\includegraphics[width=0.49\textwidth]{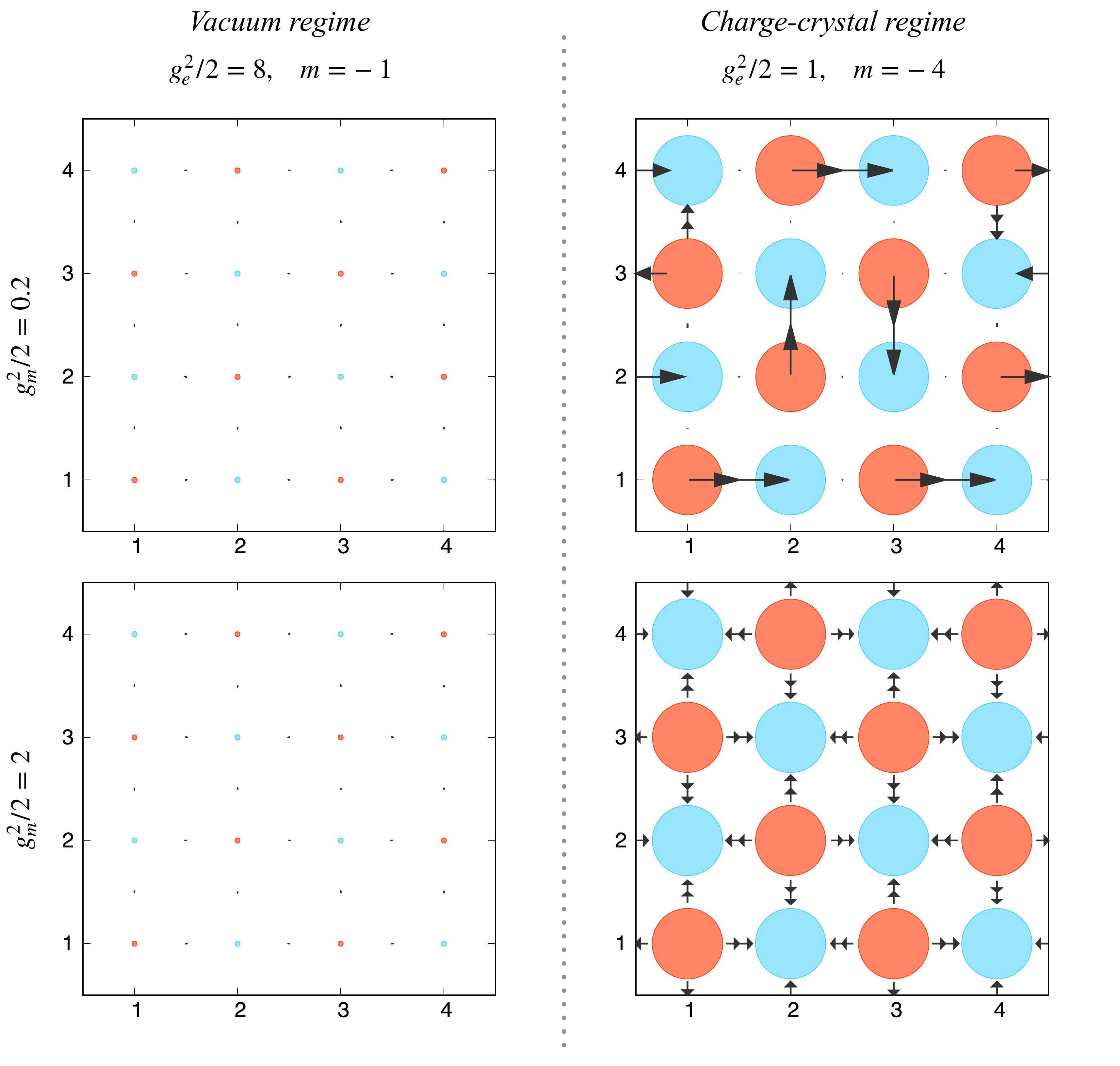}
\caption{\label{fig:field_plot_mag}
Numerical results obtained via TTN simulations.
The field plots reproduce the 
matter and gauge configurations 
for a $4 \times 4$ subsystem embedded in a $8 \times 8$ lattice with periodic boundary conditions.
Red (blue) circles represent particles (antiparticle): 
their diameter is indicating the average density, from $0$ (empty sites) to $1$ (completely filled).
The arrows in between represent the electric field: 
larger arrows indicate greater electric flux.}
\end{figure}

\begin{figure*}[t!]
\includegraphics[width=0.8\textwidth]{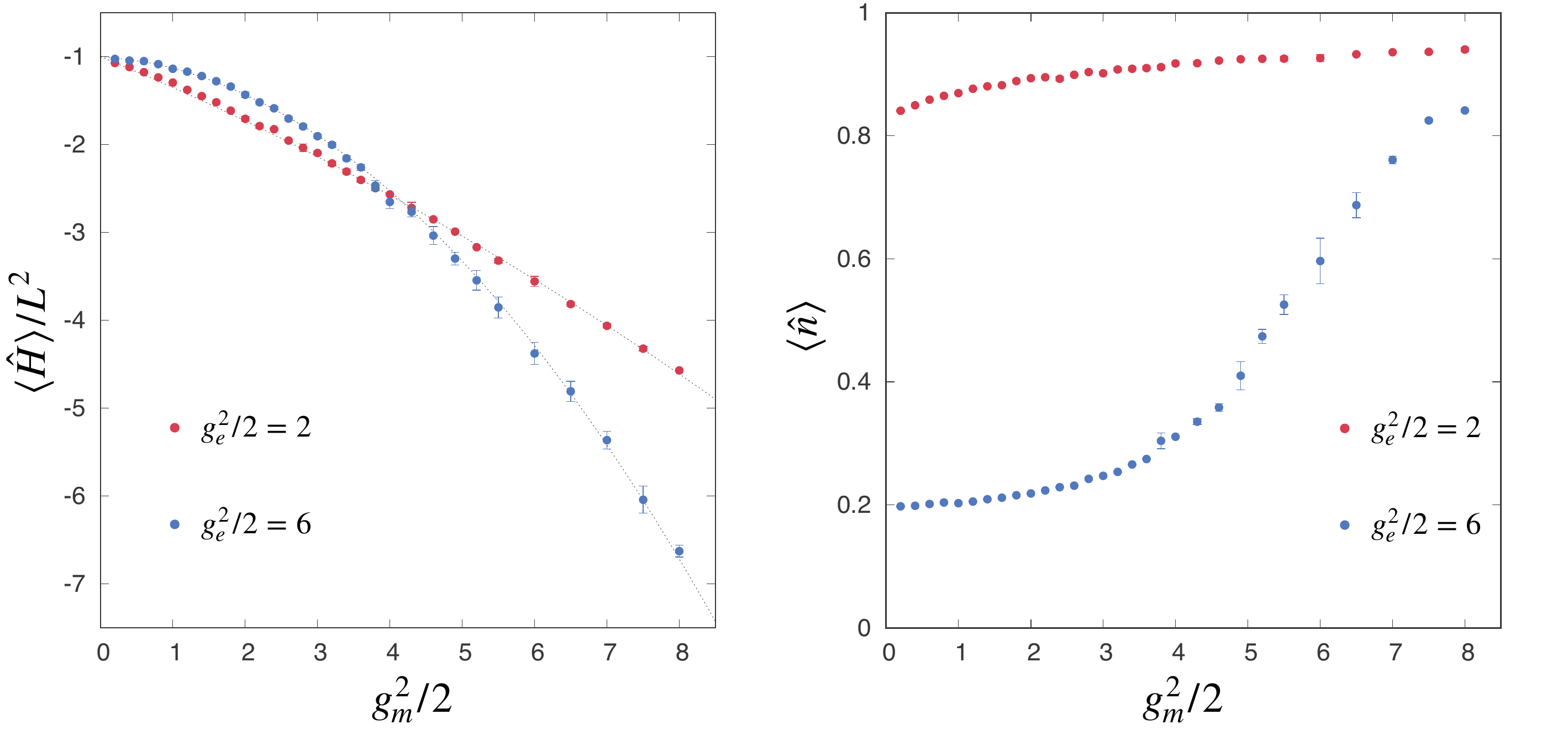}
\caption{\label{fig:magnetic_dim}\label{fig:magnetic_vac}
Numerical results for the ground state energy $\langle \hat H \rangle /L^2$ density and the particle density $\langle \hat n \rangle $
as a function of the magnetic coupling for a $8 \times 8$ systems.
The mass coupling has been fixed to $m = -2$.
The energies have been renormalised to their absolute value at $g_{m} = 0$;
dashed lines are guide for the eyes.
Data have been obtained by extrapolating from TTN simulations with different auxiliary dimensions. 
}
\end{figure*}

\subsection{Finite magnetic-coupling effects}\label{sec:mag-coupling}
We now analyze the case of non-vanishing magnetic coupling $g_m$,
especially focusing on how it impacts the many-body quantum features at zero temperature.

In Figure~\ref{fig:field_plot_mag} we show the field-plot representations
of the ground-state typical configurations for an $8 \times 8$ system 
in the presence of magnetic couplings. 
For the sake of visibility we only plot a $4 \times 4$ subsystem out of the complete $8 \times 8$ lattice simulated with periodic boundary conditions.
Both mass $m$ and electric coupling $g_e$ have been
chosen so that the system is well deep within the two different regimes
(left panels: vacuum, right panels: charge crystal).
As the magnetic coupling $g_m$ is increased to commensurate values (bottom panels)
we see negligible changes affecting the vacuum configuration.
By contrast, in the charge-crystal regime, the
non-vanishing magnetic coupling introduces a nontrivial reorganization
of the electric fields. 

Such an effect can be well understood in terms of perturbation theory: 
{\it (i)} in the vacuum region, the ground-state is not degenerate and 
the first nontrivial corrections are given by coupling such state with all the states with a single flux loop over 
a single {\it plaquette} (whose energy is therefore $2 g^{2}_{e}$).
In this regime, the flux loop state has high electric field energy,  
thus it will impact only slightly the global features of the state.   
The first order correction to the ground-state energy will 
be quadratic in the magnetic coupling, i.e.  $\sim g^{4}_{m}/g^{2}_{e}$ 
(see Fig.~\ref{fig:magnetic_vac} left panel).
Let us stress that, even though the state may experience an electric-field 
reconfiguration due to the ``field-loop'' superpositions, 
the fact that in this regime the electric field is almost zero,
causes no visible effect in its expectation value; 
this is pretty clear from the left column of Fig.~\ref{fig:field_plot_mag}:
when passing from a small ($g^{2}_{m}/2=0.2$) 
to a slightly bigger value ($g^{2}_{m}/2=2$) of the magnetic coupling,
the changes in the expectation value of the gauge field are negligible.
{\it (ii)} in the charge-crystal configuration, the effect of the magnetic interaction is nontrivial.
Indeed, the ground-state energy sector in this regime is highly degenerate, 
and the magnetic field contributes to lift such degeneracy (at much lower order than $t$).
The magnetic coupling introduces first order transitions between different gauge field configurations
therefore its first contribution to the ground-state energy is order $\sim g^{2}_{m}/g^{2}_{e}$.
Actually, a sufficiently large value of the magnetic coupling $g_{m}$ 
helps the TTN wave function to restore the square lattice symmetry
by introducing a gap between the actual ground-state and 
all the energetically unfavorable configurations.
This property is noticeable in Fig.~\ref{fig:field_plot_mag} (bottom-right panel)
where for $g^{2}_{m}/2=2$ the gauge field distribution becomes uniform
(on average) in the bulk. In this scenario, the charge crystal does not encourage the formation of dimers, but instead a global
entangled state of gauge fields.

\begin{figure*}[t!]
\includegraphics[width=0.95\textwidth]{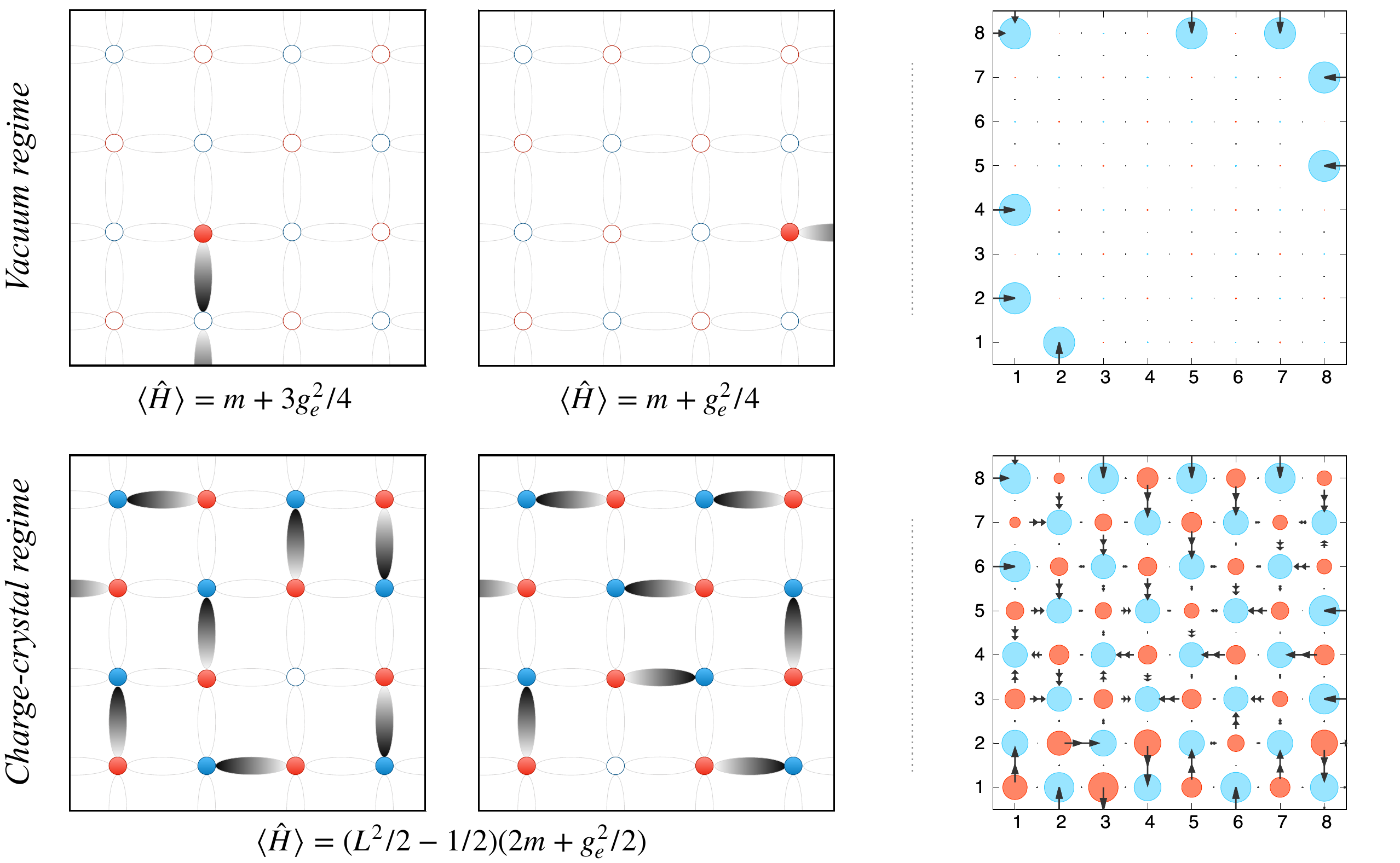}
\caption{ \label{fig:field_plot_finite_charge}
Field plots in the finite-charge density sectors: 
top row refers to the vacuum regime,
bottom row to the charge-crystal regime.
On the left of the figure, the four panels represent 
a sketch of the classical configurations (i.e. $t=0$)
for a $4\times 4$ system with open boundary conditions in the $Q=1$ charge sector. 
 The gauge field is now allowed to get out of the system by paying half-link price. 
 The panels on the right of the figure are the field plots obtained by numerical TTN simulations in $8 \times 8$
systems for the two different regimes, namely $g^{2}_{e}/2=2$ and $m = 4$ (top) or $m = -2$ (bottom),
and charge sector $Q = -8$.
In the vacuum regime, the excess of charge prefers to be localized at the boundaries,
since such configurations are more energetically favorable.  
In the dimerised regime, the holes may occupy any position since 
the system can reconfigure the pairs of dimers 
in a way to pay always the same amount of energy. 
However, due to the very high degeneracy of the low-energy sector, 
the TTN simulations may get stuck into a slightly asymmetric configuration.
}
\end{figure*}

The previous considerations are supported by 
the behavior of the ground-state energy 
and of the particle density, as a function of $g_{m}$. 
In Fig.~\ref{fig:magnetic_dim} we plot the numerical results obtained 
via TTN simulations in a $8 \times 8$ system. 
We fixed the value of the mass to $m=-2$,
and explored the behavior for $g^{2}_{e}/2 = 2$ and $6$,
that are slightly below or above the classical transition point 
$g^{2}_{e}/2 = -2m$.
We vary the magnetic coupling in a rather big interval $g^{2}_{m}/2 \in [0, 8]$.
In the first case, i.e. when the system is initially in the charge crystal configuration, 
we expect linear corrections to the ground state energy as a function of $g^{2}_{m}$;
this is pretty clear from Fig.~\ref{fig:magnetic_dim} where, 
however, some deviations are visible due to the vicinity of the phase boundary.
In the second case, i.e. starting from a quasi-vacuum, 
a quadratic deviation of the ground state is clearly visible
(see Fig.~\ref{fig:magnetic_vac}).

Interestingly, in the parameter region we are exploring, 
the magnetic coupling enhances the production of particles, 
thus increasing the average matter density,
even though the magnetic term does not directly couple to matter.
Such emergent behavior is physically relevant, since it also arises when performing phase diagram simulations
along a {\it physical line} of the QED problem. Specifically, setting $g_e \cdot g_m = 8 t^2$ realizes the physical scenario of QED.
Figure \ref{fig:physicalline} in the Appendix shows a growing charge density
at smaller QED couplings $g$, even when the (negative) bare mass is small.

In practice, the magnetic coupling creates resonating configurations of the gauge fields in the crystal charge regime,
thus decreasing the electromagnetic energy density of the state itself, which in turn favors 
the crystal charge configuration in proximity of the phase boundary. 
Hence small $g_m$ values effectively enlarge the charge crystal regime.
However, in the Spin-$1$ representation of the gauge field, 
the dimerised configuration is not stable under an arbitrary large value of the magnetic coupling,
and we expect $\langle \hat n_{x} \rangle = 0$ when $g_{m}\gg g_{e}$. 
This can be easily understood at the classical level ($t=0$)
comparing the effect of the a Wilson loop operator in the zero-matter (vacuum) sector 
and in the full-matter sector:
in the former case, each single {\it plaquette} is resonating between three different  
diagonal gauge-field configurations
$\{| \circlearrowleft \rangle,\;\vac, \; |\circlearrowright\rangle\}$, 
in the last case, only two configurations are resonating, e.g.
$\{| \uparrow\downarrow \rangle, \; |\leftrightarrows\rangle\}$,
since constructing a clockwise(anticlockwise) electric loop $\circlearrowright$ ($\circlearrowleft$)
on top of the first(second) state is forbidden by the Spin-$1$ finite representation.
This originates an energy gain which is proportional to $-\sqrt{2} g_m^2$ for a {\it plaquette} 
in the vacuum, whilst it is only $-g_m^2$ for a dimerised {\it plaquette}. In practice such difference remains at 
the many-body level as well, and, therefore, although for $-2m \gg g^2_{e}/2$ the dimerised configuration represents
the lower energy state at $g_m =0$, it will become energetically unfavorable for sufficiently strong magnetic couplings.

We have further analytical confirmation of such behavior 
from the exact diagonalisation of $2 \times 2$ systems 
(see Appendix~\ref{app:2x2} for details):
for a single {\it plaquette} system, 
the first visible effect of a non-vanishing magnetic coupling 
is to mix up the two dimerised states into two different superpositions with different energies. 
The transition between the vacuum state toward the lower energetic charge-crystal state 
is therefore sharpened and its position is shifted as well in
$
g^{2}_{e}/4+m \simeq 
(
g^{2}_{e} + g^{2}_{m}/2 - \sqrt{g^{4}_{e} + g^{4}_{m}/2}
)/4.
$
Interestingly, depending on the values of $g_{e}$,
this shifting is not monotonous in $g_{m}$,
producing an initial increase in the particle density 
followed by a definitive decrease toward zero ({\it c.f.} Fig.~\ref{fig:2x2_magnetic}),
and thus confirming the previous heuristic argument based on perturbation theory.
Again, this is a strictly finite-Spin representation effect and it does disappear as the
Spin gets larger,
as shown by analyzing the behavior for the single {\it plaquette} 
in the Spin-$2$ compact representation of the gauge field (see Appendix~\ref{app:2x2}).

\section{Finite charge density sector}\label{sec:finite_charge}

One of the most intriguing phenomena we observed in our numerical simulations rely 
on the possibility to create a charge imbalance into the system.
This scenario is challenging for Montecarlo techniques as it produces the sign problem \cite{banuls19-1,tw05}.
Instead, our gauge invariant tensor network approach is very well suited to overcome such difficulty:
the fact that the global $U(1)$ symmetry has been explicitly embedded into the 
Tensor Network ansatz~\cite{TTNA19},
allows to work exactly within each sector with fixed total charge. 
In the following we only considered $g_{m} = 0$.
Moreover, in this setup, due to the finite net electric flux coming out from the entire system, 
we have to work with open boundary conditions.
In this geometry, the dressed sites at the boundary are now characterized by one outgoing
half-link (two in the corners) which can support electric field to allow 
the existence of a non-vanishing total outgoing flux.

\begin{figure}[t!]
\includegraphics[width=0.49\textwidth]{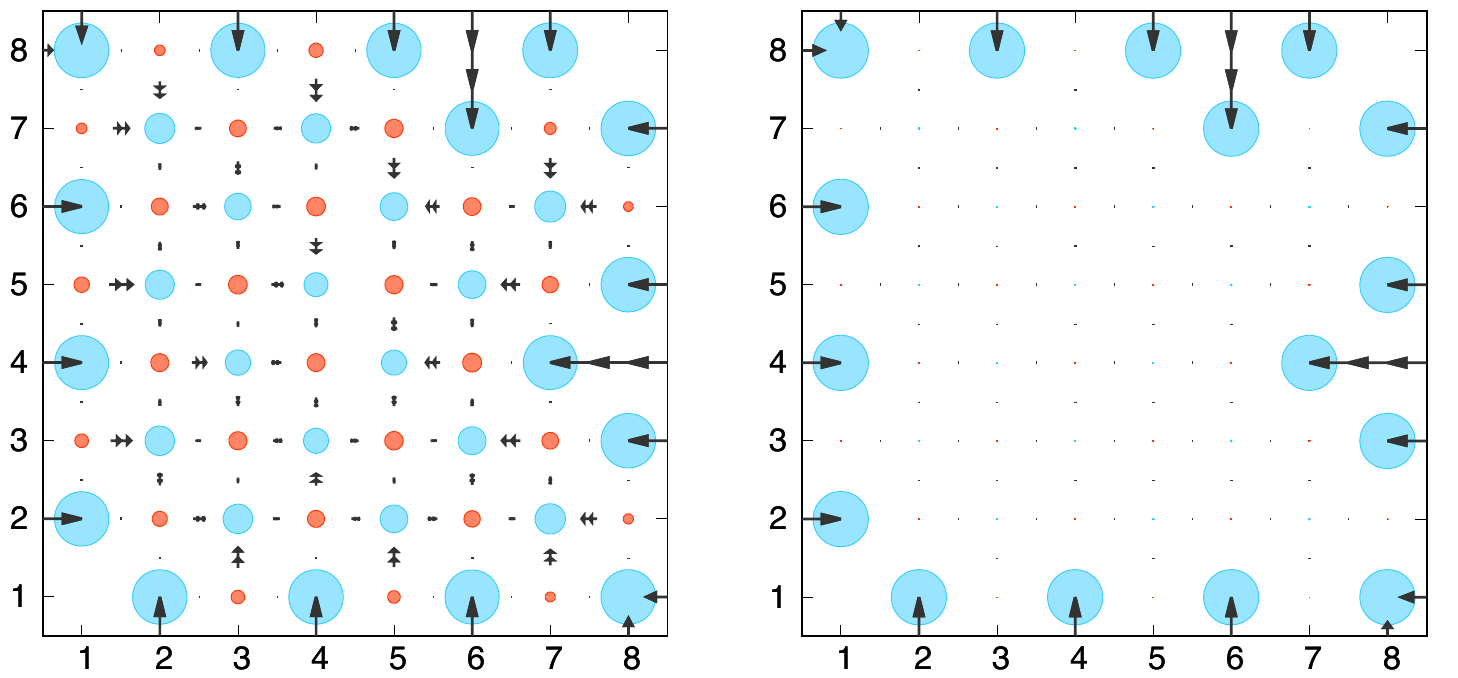}
\caption{ \label{fig:field_plot_finite_charge_Q16}
Field plots from the TTN numerical simulations of
$8 \times 8$ systems in the $Q=-16$ charge sector. 
The couplings are tuned in such a way that the system is 
deep in the vacuum regime (right panel, $g^{2}_{e}/2=2,\, m=5$),
or near the critical region (left panel, $g^{2}_{e}/2=2,\, m=-1$).
Notice how the excess of charge is larger than the allowed 
antimatter sites at the surface: the system has to allocate 
two extra charges in the next-neighboring sites to the surface.
}
\end{figure}

When a finite density of charge $\rho \equiv \langle \hat  Q\rangle /L^{2} \in \{-1/2,1/2\}$
is injected into the system, we expect a different behavior
depending on the part of the phase diagram the ground state is belonging within.
Indeed, when the ground state is very close to the bare vacuum,
any charge created on top of it is forced to reach the boundaries so as to minimize the total energy; 
this is easily understood already with the classical ($t=0$) Hamiltonian, 
and there are no fluctuations of the gauge fields. In this case,  
a classical configuration with a single charge located at distance $\ell$
from the boundary costs at least $\ell g^{2}/2$ more than the optimal 
configuration where the same charge is located at the surface  (see Fig.~\ref{fig:field_plot_finite_charge}). 
In this regime the diagonal energy term gets modified as $E_{v}/L^{2} = (g^{2}_{e}/4 + m)\rho$, 
as far as $\langle \hat  Q\rangle \leq 2(L-1)$,
i.e. whenever the total excess of charge is lower than the number of allowed free sites 
at the boundaries.
When the total charge gets larger, deeper sites start to be filled, e.g.
for $2(L-1) < \langle \hat  Q\rangle \leq  4(L-2)$ one starts filling 
the next-neighboring sites to the surface (e.g. Fig.~\ref{fig:field_plot_finite_charge_Q16}).
Overall, this argument support the existence of a {\it phase separation} 
between a boundary region attached to the surface, or strip,
where charges aggregate, and a bulk
region expelling charges and electric fields.
In the picture where the gauge field is not truncated ($S \to \infty$), both regions will scale as a surface.
In practice, defining $\rho_{\ell} \equiv 2 \ell(L-\ell)/L^2$ 
being the maximum amount of charge-density the system can store within a strip 
of extension $\ell$ from the surface, we have a sharp discontinuity in local charge densities,
at the smallest $\ell^{*}$ such that  $\rho \leq \rho_{\ell^{*}}$, between a finite-charge region (for $j < \ell^{*}$)
and a zero-charge region (for $j>\ell^{*}$). 
In particular, in the thermodynamic limit we obtain $\ell^{*}/L = (1-\sqrt{1-2|\rho|})/2$.
In other words, the width $\ell^{*}$ 
of the surface strip where all charges are localized,
varies smoothly in $[0,L/2]$ as $|\rho|$ varies in $[0,1/2]$.
Quantitatively, both the depth of the surface strip ($\ell^{*}$), 
and the diameter of the bulk region ($L/2 - \ell^{*}$) scale linearly with $L$,
thus the phase separation argument applies when approaching the thermodynamic limit, as long as
the gauge field is unconstrained and the lattice spacing stays finite.
in practice, as far as the average charge density is finite,
we always have an extensive region in the bulk of the system, 
whose linear dimension scales as 
$
L\sqrt{1-2|\rho|}/2
$, 
which exhibits no charges.
We stress, however, that when introducing a fixed truncation of the gauge field (to any spin $S$) the amount of total charge that can be injected in
the system is limited to a linear scaling in $L$ since, due to the Gauss' law, the total electric flux at the boundary must match the total charge.
This implies that to approach the thermodynamical limit at finite charge density one needs to increase the truncation or introduce static background fields.


We expect this picture to be slightly modified at finite hopping coupling $|t|$,
but to remain valid as far as the system belongs to the vacuum regime. In practice,
a finite tunneling amplitude introduce a small homogeneous particle density, thus 
slightly rising $\ell^{*}$ and having the effect as well to 
build up a finite {\it charge-penetration length}
$\sim |t|$ such that the transition at $\sim \ell^{*}$ becomes smooth, 
with an exponentially small density-charge tail penetrating into the bulk.
The overall scenario is confirmed by the field plots in Fig. \ref{fig:field_plot_finite_charge} 
and Fig.~\ref{fig:field_plot_finite_charge_Q16},
where it is pretty clear that when the couplings are tuned in order for the system to be deep into the vacuum regime,
the excess of charges stick to the boundary so as to minimize the length of the attached electric strings. 
In principle, all possible configurations with all charges at the boundaries are energetically equivalent.
Anyhow, the TTN many-body wave function spontaneously breaks such symmetry and 
picks up a single specific configuration, as it is usually the case with DMRG-like algorithms.
We stress that such phase separation, where both bulk and boundary regions scale extensively,
is likely an artifact of the lattice discretization, where the amount of local charge density is bound.

When the state belongs to the charge crystal regime,
a finite positive (negative) charge density is mainly generated by creating holes
in the odd (even) sub-lattice; namely, negative (positive) charges are removed from the fully dimerised state.
In order to minimize the energy, the holes can be now fully delocalised:
a hole in the bulk, or at the boundary, generates a reconfiguration of the charge crystal 
state in such a way to always pay the same amount of energy,
and guarantee the expected total outgoing electric flux. 
In this regime, the zero-order energy term gets modified as $E_{d}/L^{2} = (g^{2}_{e}/4 + m)(1-\rho)$.
The entire system is now characterized by a unique spatial phase where we expect 
a uniform average charge density and finite electric field in the bulk. 
Let us mention that, for any finite value of the hopping amplitude, we still expect a similar behavior,
where the transition toward the phase-separated phase will be driven by the competition
between the mass and the electric coupling. 

\begin{figure}[t!]
\includegraphics[width=0.5\textwidth]{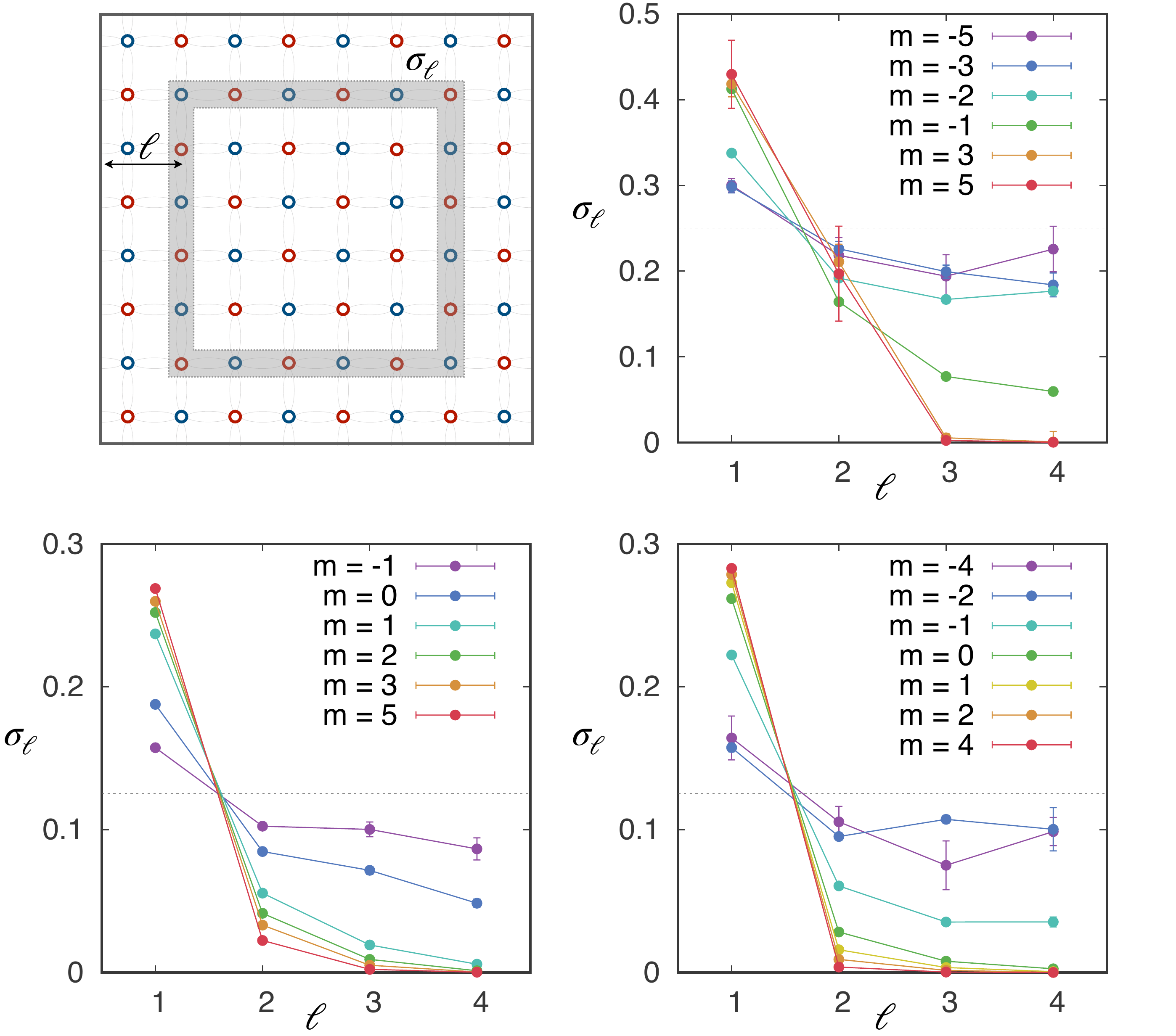}
\caption{ \label{fig:surface_charge}
Surface charge density evaluated in $8 \times 8$ system as sketched in the top-left image:
the shaded region represents the domain $\mathcal{D}_{\ell}$ defined in the main text.
TTN simulations have been performed for different charge sectors and electric couplings;
in clockwise order: $(Q=-16,\, g^{2}_{e}/2=2)$,  $(Q=-8,\, g^{2}_{e}/2=2)$ and  $(Q=-8,\, g^{2}_{e}/2=1/2)$.
}
\end{figure}

\begin{figure*}[t!]
\includegraphics[width=0.7\textwidth]{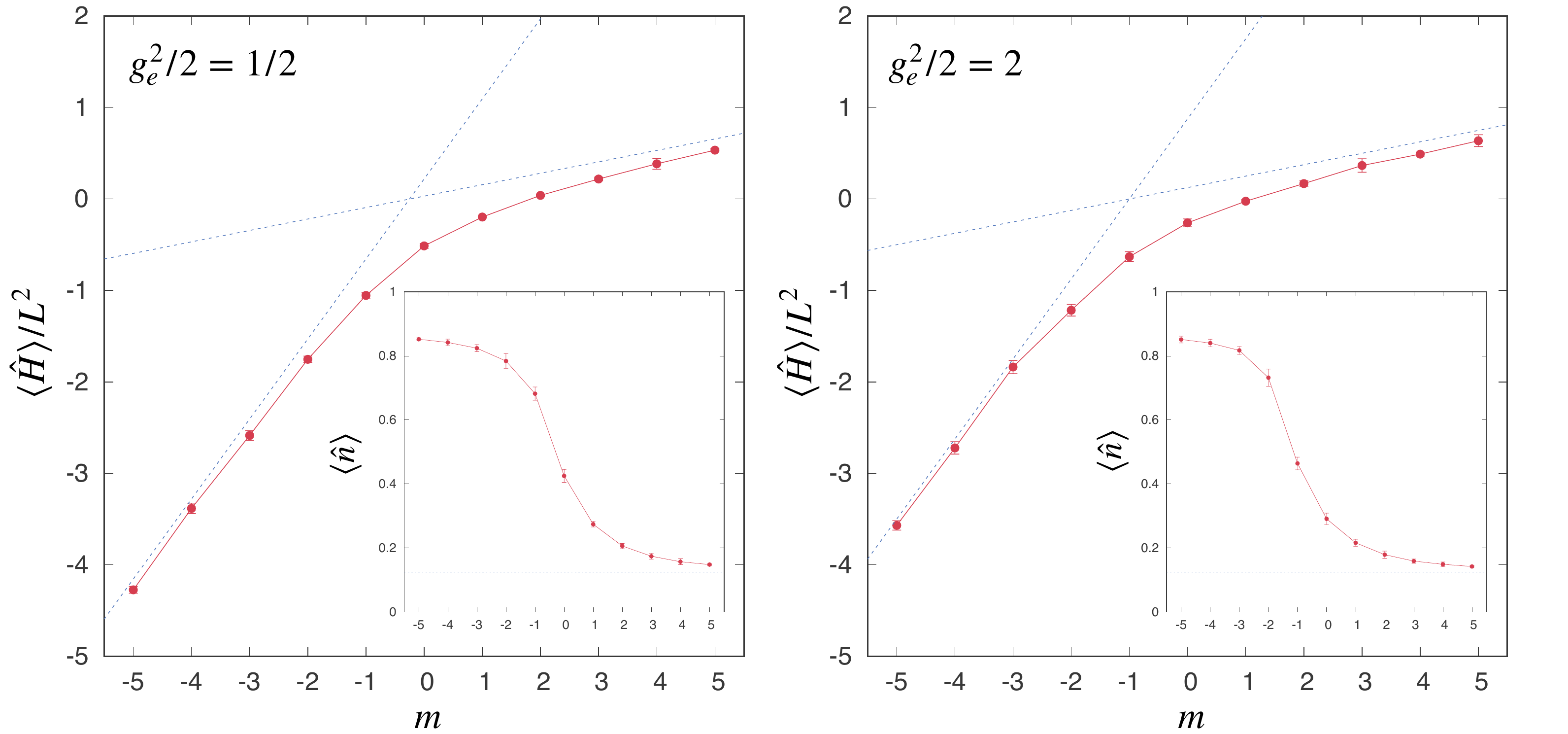}
\caption{ \label{fig:energy_particle_finite_charge}
Ground-state energy density and particle density (insets) as a function of the bare mass $m$ for $8 \times 8$ lattice with $\rho = -1/8$, i.e. in the $Q=-8$ charge sector. 
In the left panel, the transition between the two regimes occurs at $m\sim -1/4$, 
while in the right panel it is located at $m\sim -1$.
Dashed lines are the asymptotic values in the zero-order perturbative approximation. 
}
\end{figure*}

In order to highlight the different features of the low-energy state
at finite chemical potential, we analyzed the behavior of the surface charge density
\be
\sigma_{\ell} \equiv \frac{1}{{\rm dim}{\mathcal{D}_{\ell}}}\sum_{x \in \mathcal{D}_{\ell}} \langle  \hat \psi^{\dag}_{x} \hat \psi_{x}\rangle
\ee
where $\mathcal{D}_{\ell}$ is a square  
which counts ${\rm dim}\,  {\mathcal{D}_{\ell}} = 4(L+1-2\ell)$ lattice sites as sketched in Fig.~\ref{fig:surface_charge}. 
Here $\ell \in \{1,2,\dots, L/2\}$ represents the distance of the domain $\mathcal{D}_{\ell}$ from the external
surface: namely, as $\ell$ grows, we select domains deeper into the bulk.  

In Fig.~\ref{fig:surface_charge} we plot the surface charge density $\sigma_{\ell}$
as function of $\ell$ for different point in the coupling-parameter space.
As far as the Hamiltonian is tuned into the vacuum regime, the surface charge suddenly drops
when getting into the bulk of the system. As expected, for finite value of the couplings,
when approaching the critical region, the bulk charge density gets enhanced;
finally, once the system reaches the charge-crystal regime, $\sigma_{\ell}$
acquires a loosely uniform shape.
 
Finally, we carefully checked the ground-state energy density and the particle density, 
which are plotted in Fig.~\ref{fig:energy_particle_finite_charge}
as a function of the mass for two different values of the electric coupling. 
Notice how, for sufficiently large positive (negative) value of the mass the data get closer to the 
perturbative predictions. The intermediate region, at $m \sim g^{2}_{e}/4$, 
is characterized by stronger quantum fluctuations and thus exhibit a smooth transition between
uniform and nonuniform charge distribution in space.

As a concluding remark we stress that, while in this section we considered the boundary conditions to be completely free,
setting a specific set of boundary conditions for the problem of electrodynamics comes with no conceptual or numerical difficulty.
Typical boundary conditions are realized by means of a boundary Hamiltonian $H_b$ to be
added to the bulk Hamiltonian $H$ from Eq.~\eqref{eq:H_QED} (see Appendix \ref{app:boundary} for details). 


\section{Conclusions}\label{sec:conclusion}

In this work we demonstrated a novel efficient tensor network approach to the study of two-dimensional lattice gauge theories. 
By exploiting the quantum link formulation of LGT, the fermionic rishon representation of quantum links, and unconstrained 
Tree Tensor Networks, we investigated the equilibrium properties of 
a two-dimensional lattice QED within its first compact spin representation. We present results for lattice size up to 16x16, whose Hilbert space dimension is approximately equivalent to that of a system composed by spins one-half on a square lattices with edges of about eighty lattice sites. 
Whenever possible, we confirmed our results with perturbative analysis and small scale exact simulations. 

In particular,  we identified different regimes at zero chemical potential, a vacuum state and a charge-density one, that reproduce what has been found in the one-dimensional case, and investigated the effects of a magnetic term uniquely present in two-dimensions. Finally, we explore the finite density scenario and individuate 
two distinct behaviors correspondent to the vacuum and charge-density configurations: in the former case, the excess charges accumulate on the boundaries to minimize 
the electric fields to be energetically sustained, as for classical charged conductors. In the latter, the excess charge is distributed uniformly in the bulk and boundaries. 

In conclusions, we have shown that unconstrained Tree Tensor Network are a powerful tool 
to obtain a non-perturbative description of a lattice gauge theory in two dimensions. We stress that these simulations have been obtained 
on standard clusters without exploiting heavy parallelization and with simulations lasting only a few days. 
Despite the fact that the presented results are not yet at the level to allow physical predictions in the continuous limit for the system we study, we foresee that upgrading the current software to exploit the full power of High Performance Computing -- without mayor changes in the algorithms -- larger system sizes, an additional dimension, the continuum and large-$S$ limit, and more complex Abelian and non-Abelian lattice gauge theories will be in range of the approach presented here, as already shown for the one-dimensional case~\cite{bccj13,pstm19}.
Our proposed architecture is perfectly tailored to accommodate advanced strategies of diagnostics, including elaborate string order parameters capable
of detecting deconfined phases and topological order \cite{Horseshoe1,Horseshoe2,HorseshoeErez}.

\section{Acknowledgments}
We are very grateful to M. Dalmonte, K. Jansen, L. Salasnich, U.J. Wiese and P. Zoller for valuable comments.
Authors kindly acknowledge support from the BMBF and EU-Quantera via QTFLAG and QTHEP, 
the Quantum Flagship via PASQuanS,
the INFN, the MIUR via the Italian PRIN2017, the DFG via the TWITTER project,
the US Air Force Office of Scientific Research (AFOSR) via IOE grant FA9550-19-1-7044 LASCEM,
and
the Austrian Research Promotion Agency (FFG) via QFTE project AutomatiQ.

\clearpage

\appendix

\section{Physical QED-scenario}

\begin{figure}[t!]
\includegraphics[width=0.5\textwidth]{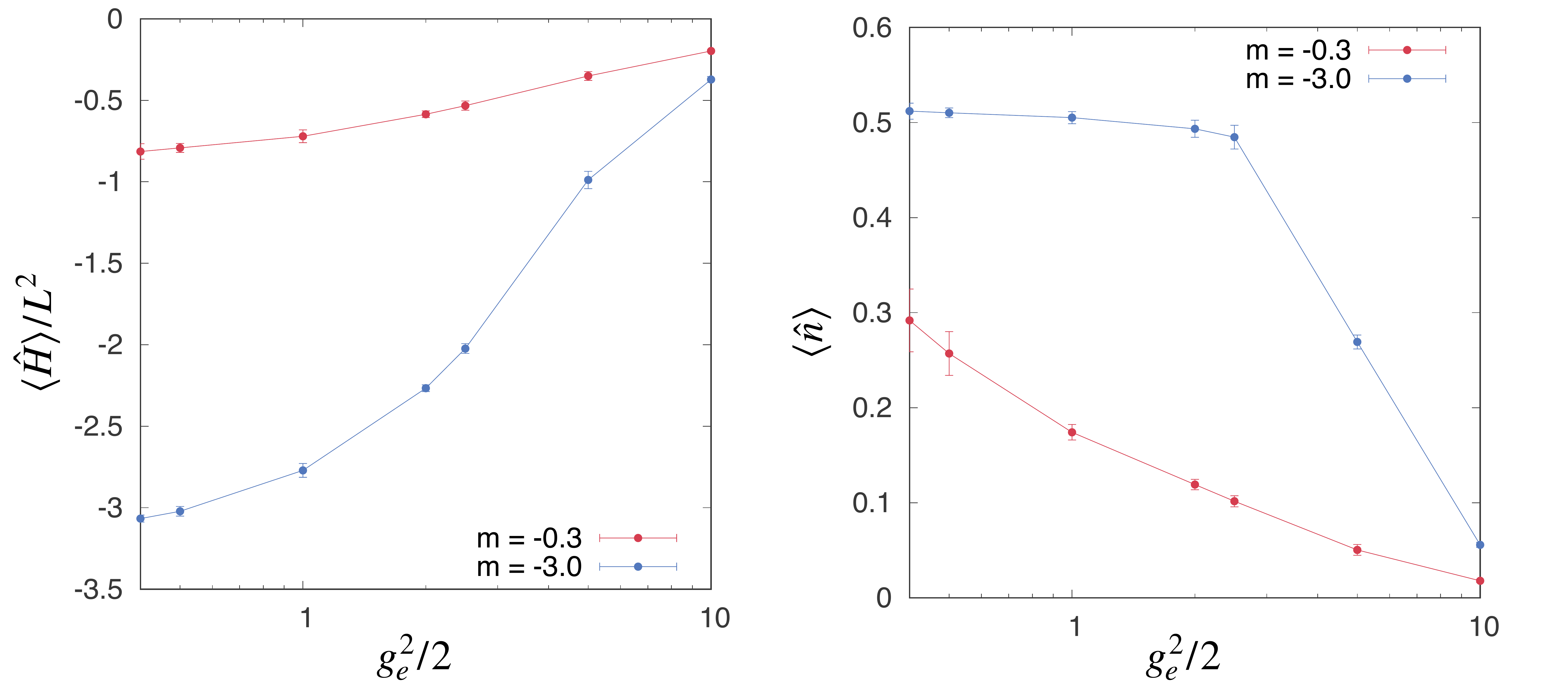}
\caption{ \label{fig:physicalline}
Numerical results for the ground state energy density $\langle \hat{H}\rangle / L^2$ and the particle density $\langle \hat{n} \rangle$ of the original QED Hamiltonian formulation as a function of the electric coupling $g_e$ for $8 \times 8$ systems. The lattice spacing is set to $a= 1/t = 1$ and the magnetic coupling is tuned with respect to $g_m^2=8/(g^2_e a^2)$. The mass coupling has been set to $m_0 = \{-0.3, -3.0\}$ respectively. The data has been obtained by extrapolating from TTN simulations with different auxiliary dimensions.
}
\end{figure}

Complementing the discussions in Sec.~\ref{sec:mag-coupling}, we present two {\it physical lines} of phase diagram simulations of the QED problem with $g_e \cdot g_m = 8 t^2$. Fig.~\ref{fig:physicalline} shows a growing charge density at smaller QED couplings $g$, even when the (negative) bare mass is small.

\section{Charge Screening}\label{app:screening}

\begin{figure}[t!]
\includegraphics[width=0.475\textwidth]{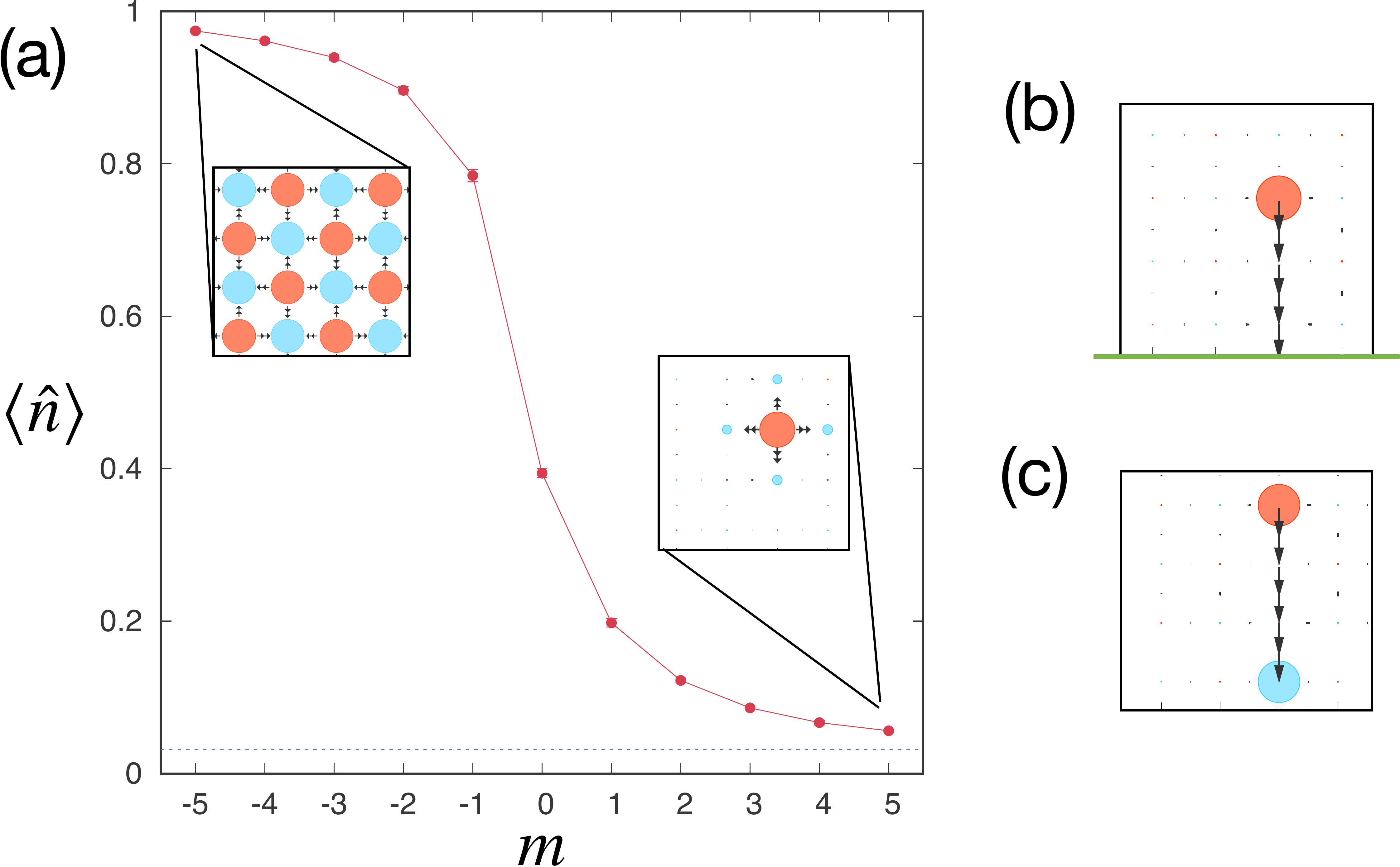}
\caption{ \label{fig:charge_screening}
Occupation for pinned charges in an 8x8 system of the original QED Hamiltonian with $g_e=1$ and $a=1/t=1$. (a) particle density for one charge pinned in the zero charge sector $Q_{tot}=0$ with respect to the bare mass $m$ for periodic boundaries. The system transitions from the completely filled charge-crystal phase to the pinned charge screening. (b) Field plot for a system with total charge $Q_{tot}=1$ with open boundaries (green line). (c) Field plot for two pinned charges in the $Q_{tot}=0$ symmetry sector. All of the shown field plots are 4x4 subsystems embedded in an 8x8 simulation. For (b) and (c) the mass term is set to $m=4$.
}
\end{figure}

Here we briefly address the problem of detecting confinement. A natural way of exhibiting confinement in our lattice scenario is to show that
electric field lines do not extend over infinite lengths in the full-fledged theory (where $g_e g_m = 2 \sqrt{2} t$), even when we enforce the presence of charges at specific locations.
In this sense, the ground QED solution adjusts the mobile charges (and anticharges) to screen the {pinned} ones.
We can insert such pinned charges by tuning a local chemical potential term $\tilde{m}_x (-1)^x \hat\psi^{\dag}_{x}\hat\psi_{x}$ which shifts the mass at site $x$ to strongly negative values
($\tilde{m}_x + m \ll -1$),
thus favoring the presence of a full charge at $x$ in the ground state.

Adding a single pinned charge has the effect of creating a local excitation in the vacuum-like regime. Figure \ref{fig:charge_screening}(a) shows a scenario with global zero charge, and a single pinned
charge, under periodic boundaries. While the crystal-charge regime is mostly unaffected, in the vacuum-like regime opposite-sign charges are attracted around the pinned one, as to form an almost perfect meson, carrying an electric charge quadrupole. Field lines propagating from this configuration are very short ranged, thus supporting confinement.

It is important to mention that confinement can be further corroborated by string-breaking analysis, where an initial (high-energy) configuration with a long field-line string breaks down to multiple localized mesons.
Configurations with long (extensive) field lines can be engineered either by field-linking a bulk charge to a point in the boundary, in the sector $Q_{tot}=1$, or by setting two pinned charges far apart in the $Q_{tot}=0$ sector. When the field lines (strings) scale in length with the system size, we expect them to be broken by the appearance of screening charges around the pinned ones, in the thermodynamical
limit. By contrast, for finite-length strings, it is possible to set the bare mass sufficiently large that they will remain unbroken, as shown in Fig.~\ref{fig:charge_screening}(b) and (c).

\section{Constructing the computational Hamiltonian}\label{app:matrix}

\begin{flushright}

\end{flushright}
In this section we sketch the steps needed to obtain the operator matrices, and their elements, which
appear in the computational formulation of the quantum link QED model. In particular, we stress how
to construct building-block operators $A_{j}^{(\alpha)}$, each acting on a single dressed site $j$, which are {genuinely local},
in the sense that they commute, by construction, with every other building-block operator at another site:
$[A_{j}^{(\alpha)},A_{j' \neq j}^{(\alpha')}] = 0$. The electric field term and the bare mass term are diagonal in the occupation
basis of fermions and rishons, as Eq.~\eqref{eq:dressed-state}, and thus trivially obtained. The non-diagonal terms
are decomposed as follows:

{\bf Matter-Field coupling terms} $-$ Matter-field terms decompose naturally as
$\psi^{\dagger}_{x} U_{x,x+\mu_x} \psi_{x+\mu_x} = A_x^{(1) \dagger} A_{x+\mu_x}^{(3)}$ (and its hermitian conjugate) for horizontal
`hopping' terms, and $\psi^{\dagger}_{x} U_{x,x+\mu_y} \psi_{x+\mu_y} = A_x^{(2) \dagger} A_{x+\mu_y}^{(4)}$ for vertical hopping.
The decomposition into building blocks is based upon
   \begin{multline}
    \psi^{\dagger}_{x} U_{x,x+\mu_x} \psi_{x+\mu_x} =
    \psi^{\dagger}_{x} \eta_{x,\mu_x} \eta^{\dagger}_{x+\mu_x,-\mu_x} \psi_{x+\mu_x}
    \\
     = ( \eta^{\dagger}_{x,\mu_x} \psi_{x})^{\dagger} ( \eta^{\dagger}_{x+\mu_x,-\mu_x} \psi_{x+\mu_x}  )
    = A_x^{(1) \dagger} A_{x+\mu_x}^{(3)},
   \end{multline}
Where $\eta_{x,\mu}$ are the 3-hardcore fermionic operators defined in Eq.~\eqref{eq:etadef}.
Both $A_x^{(1)}$ and  $A_x^{(3)}$ are built on an even number of fermionic operators, thus they commute with
any operator which does not act on site $x$, thus genuinely local.
The vertical hopping term is similarly decomposed into building-block operators.

{\bf Magnetic terms} $-$ The magnetic (or plaquette) term decomposes into building block operators, acting on the four dressed sites at
the corners of a plaquette. Specifically, we have
\begin{multline}
  U_{x,x+\mu_x} U_{x+\mu_x, x+\mu_x+\mu_y} U^{\dagger}_{x+\mu_y, x+\mu_x+\mu_y} U^{\dagger}_{x,x+\mu_y}
  = \\ =
  \eta_{x,\mu_x} \eta^{\dagger}_{x+\mu_x, -\mu_x}
  \eta_{x+\mu_x, \mu_y} \eta^{\dagger}_{x+\mu_x+\mu_y, -\mu_y}
  \\ \times
 \left(  \eta_{x+\mu_y,\mu_x} \eta^{\dagger}_{x+\mu_x+\mu_y, -\mu_x} \right)^{\dagger}
 \left(  \eta_{x,\mu_y} \eta^{\dagger}_{x+\mu_y, -\mu_y} \right)^{\dagger}
 \\ = 
  - \left( \eta^{\dagger}_{x,\mu_y} 
  \eta_{x,\mu_x} \right)
  \left( \eta^{\dagger}_{x+\mu_x, -\mu_x}
  \eta_{x+\mu_x, \mu_y} \right) 
  \\ \times
  \left( \eta^{\dagger}_{x+\mu_x+\mu_y, -\mu_y}
 \eta_{x+\mu_x+\mu_y, -\mu_x} \right)
 \left( \eta^{\dagger}_{x+\mu_y,\mu_x}
 \eta_{x+\mu_y, -\mu_y}  \right)
 \\ \equiv
 - A_x^{(5)} A_{x+\mu_x}^{(6)} A_{x+\mu_x+\mu_y}^{(7)} A_{x+\mu_y}^{(8)},
 \end{multline}
to be added, in the Hamiltonian, to its Hermitian conjugate.
All operators in this decomposition are local and ready to use for TTN algorithms.



\section{Tensor Networks}


\begin{figure}
\centering
\begin{minipage}{.5\linewidth}
  \centering
  \includegraphics[width=0.9\linewidth]{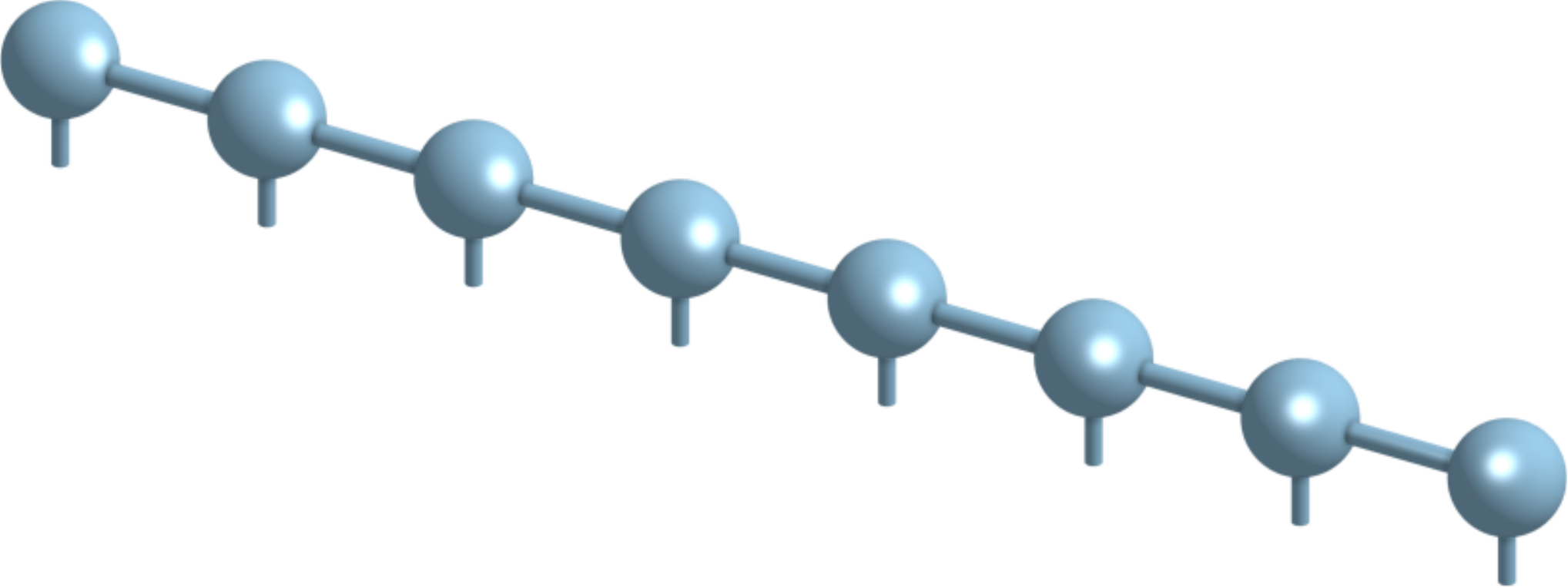}
  \subcaption{Matrix Product States}
  \label{fig:Tagging}
\end{minipage}%
\begin{minipage}{.5\linewidth}
  \centering
  \includegraphics[width=0.9\linewidth]{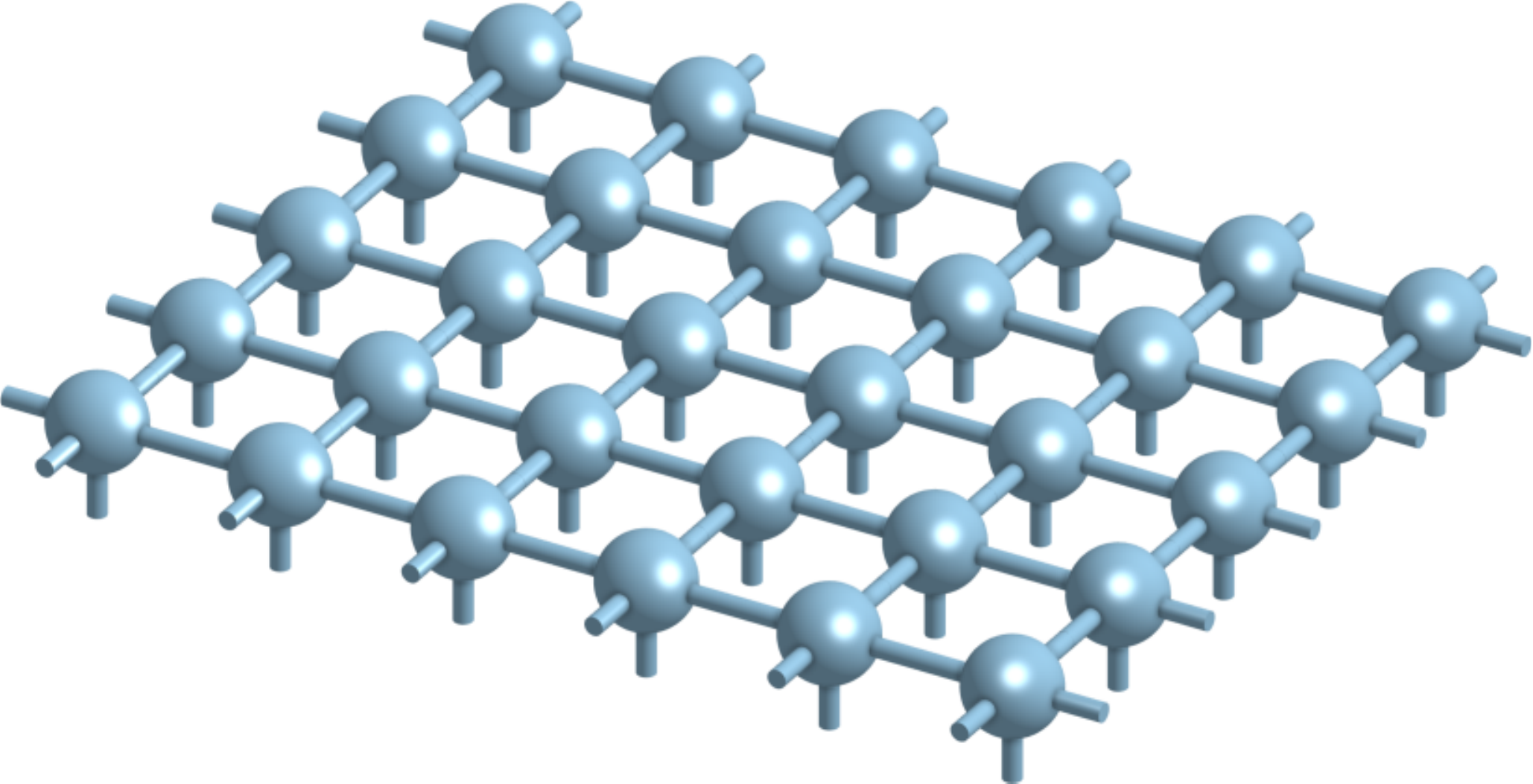}
  \subcaption{Projected Entangled Pair States}
  \label{fig:probCorr}
\end{minipage}
\begin{minipage}{.5\linewidth}
  \centering
  \includegraphics[width=0.9\linewidth]{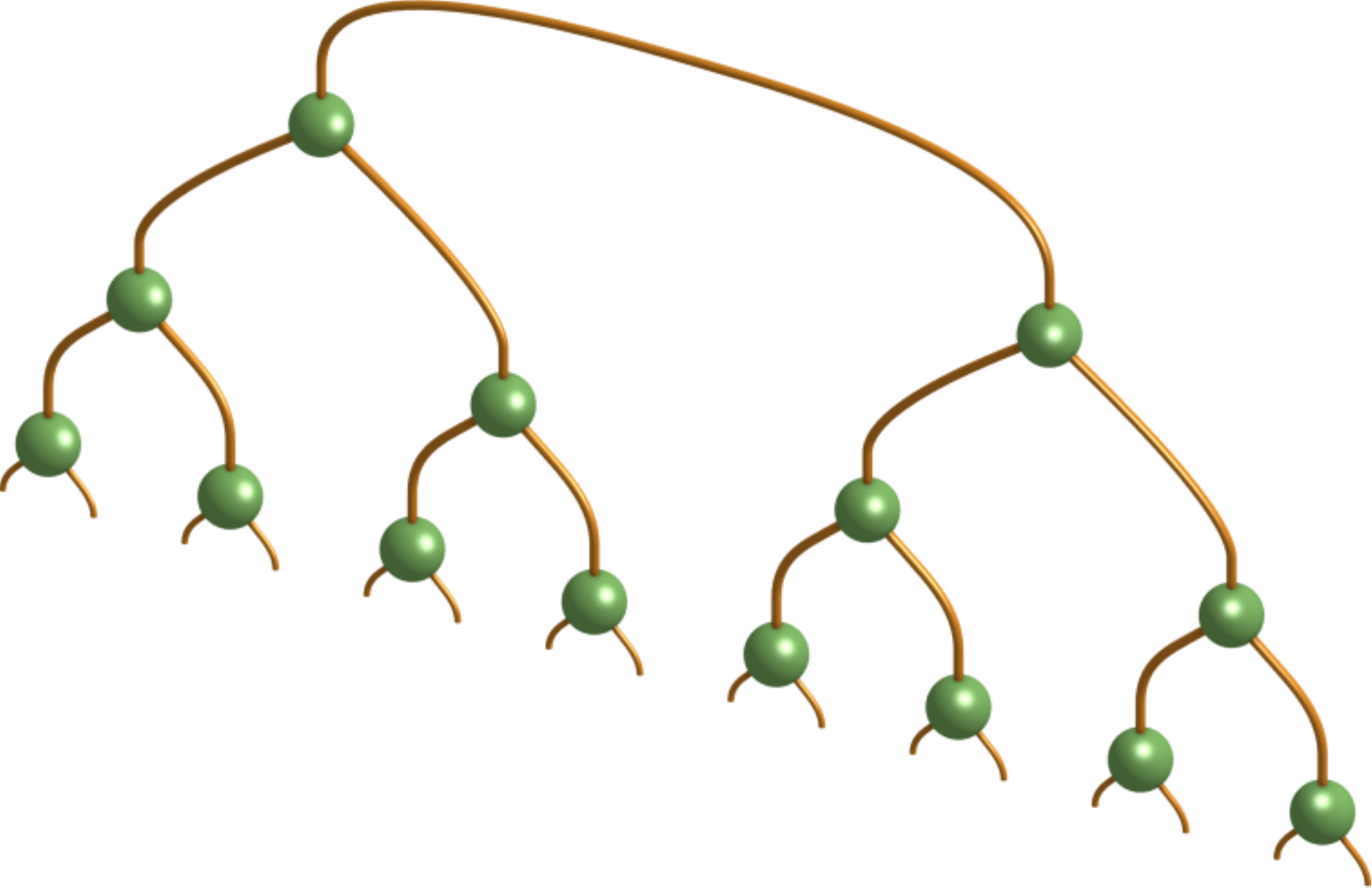}
  \subcaption{Tree Tensor Networks}
  \label{fig:probDNN}
\end{minipage}%
\begin{minipage}{.5\linewidth}
  \centering
  \includegraphics[width=0.6\linewidth]{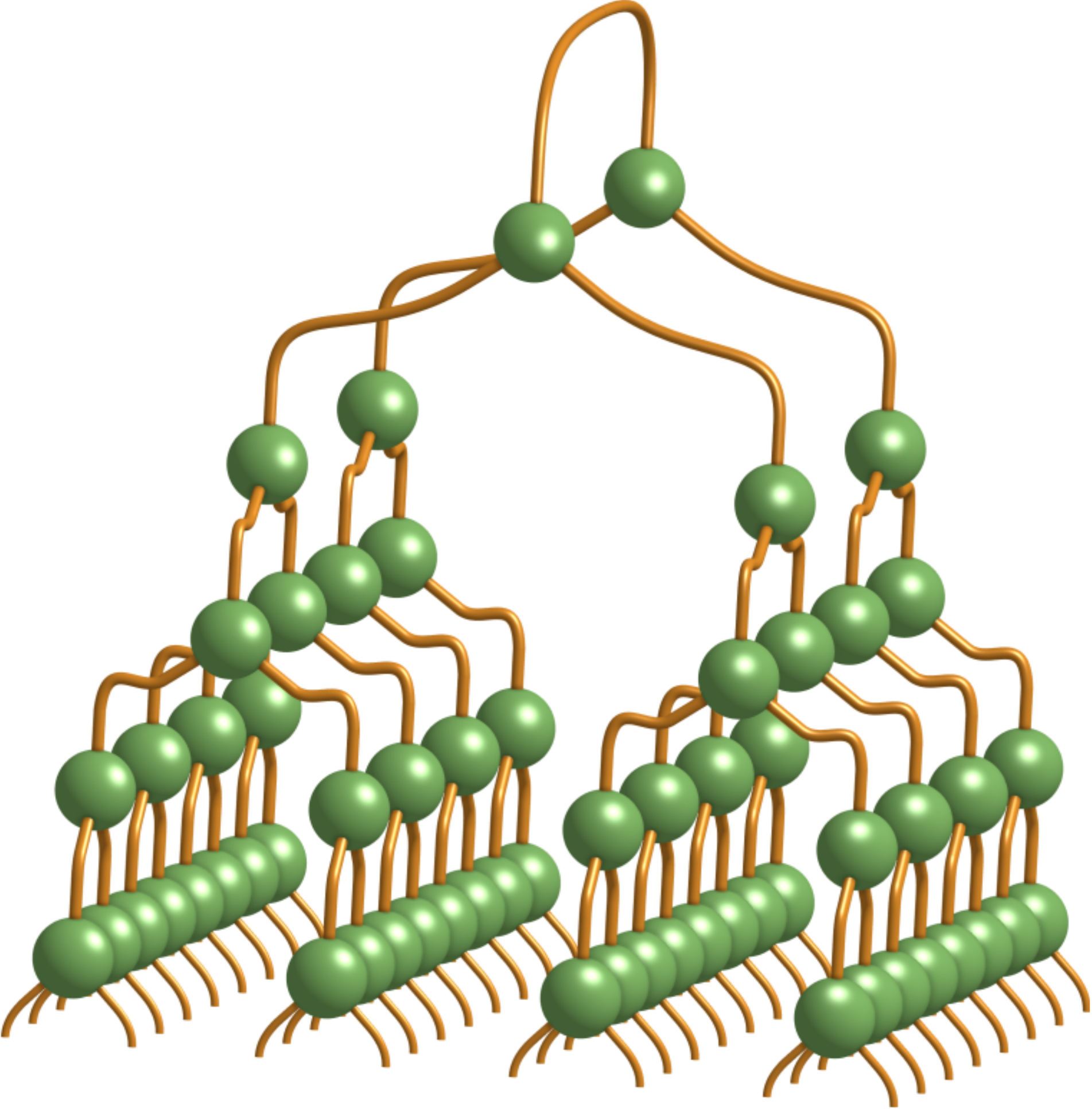}
  \subcaption{Tree Tensor Networks}
  \label{fig:probTTN}
\end{minipage}
\caption{
Tensor Network representations for a quantum many-body wavefunction: The MPS and TTN for 1D systems (left), and the PEPS and TTN for 2D systems (right).}
\end{figure}

In what follows, we describe the background and main principles of Tensor Networks and, in particular, the TTN ansatz considered in this work. For a more in-depth description of TNs, we refer to more technical reviews and text books \cite{TTNA19,OrusReview,montangero18,DMRGageofMPS}.

TNs are used to efficiently represent (pure) quantum many-body wavefunctions $|\psi\rangle$, which live in the tensor product $\mathcal{H} = \mathcal{H}_1\otimes \mathcal{H}_2\otimes \cdots \mathcal{H}_N$ of $N$ local Hilbert spaces $\mathcal{H}_k$, each assumed to be of finite dimension $d$. Expressing such a state in real-space product basis means decomposing the wavefunction as
\begin{equation}
|\psi \rangle = \sum_{i_1,...i_L = 1}^{d}{c_{i_1,...,i_L} |i_1\rangle_1 \otimes |i_2\rangle_2 \otimes ...\otimes |i_L\rangle_L} ~,
\label{TN-eq:totalstate}
\end{equation}
where $\{ |i\rangle_k \}_i$ is the canonical basis of site $k$, spanning $\mathcal{H}_k$.
Describing such a general state by all possible combinations of local states requires $d^N$ coefficients $c_{i_1,...,i_N}$. Thus, we have exponential growth with the system size $N$ in the exact representation of the wave-function.
For {\it physical} states, which satisfy certain entanglement bounds under real-space bipartitions (area laws) \cite{PEPSbasics2006,MPSfaithful}, Tensor Networks offer a more efficient representation. This is done by decomposing the complete rank-$N$ tensor into a set of local tensors with smaller rank, connected with auxiliary indices. We control the dimension of the auxiliary indices with the bond-dimension $\chi$ and thereby the amount of captured information. Thus, tuning this parameter $\chi$, TNs interpolate between a product state, where quantum correlations are neglected, and the exact, but inefficient representation. The most prominent TN representations are the Matrix Product States (MPS) for 1D systems \cite{OstlundRommer,MPSfaithful,DMRGageofMPS}, and their higher-dimension variant: the Projected Entangled Pair States (PEPS) \cite{PEPSfirsttime2004, PEPSbasics2006,OrusPEPS}, while Tree Tensor Networks (TTN)~\cite{VidalDuanTTN2006,HomogeneousTTN,TTN14,gqh17} (as well as MERA \cite{MERAfirst,MERA2D}) can in principle be defined in any lattice dimension.

Algorithms for MPS have been developed for over 20 years, and have established the MPS ansatz as primary workhorse for equilibrium problems in 1D \cite{WhiteDMRG92,DMRGageofMPS}, and in many cases even out-of-equilibrium \cite{TDMRG,TEBD}.
By contrast, the development of TN algorithms, which are both quantitatively accurate and polynomially scalable, for two-dimensional lattices is still ongoing.
Currently, we are still facing the open question of which Tensor Network geometry is generally best suited for 2D simulations. The PEPS approximates the complete rank-$N$ tensor by a decomposition with one tensor for each physical site. These tensors are then connected through a grid analogous to the lattice, resulting in a TN with `loops' (non-local gauge redundancies). On the other hand, TTNs represent the wavefunction with a network geometry without loops, thus allowing (polynomially-scaling) universal contraction schemes \cite{TTNA19}.

By its structure, the PEPS is the intuitive (and potentially more powerful) representation of a 2-dimensional quantum many-body wavefunction satisfying the area laws of entanglement. However, in general, it lacks an exact calculation of expectation values. In fact, for a finite square lattice with $N= L\times L$ sites, the contraction of the complete PEPS to perform this calculation scales exponentially on average system length $L$~\cite{JensPEPSarehard}. Additionally, the optimization of the PEPS ansatz has a higher numerical complexity $\mathcal{O}(\chi^{10})$ with the bond-dimension, so that the typical bond-dimensions achieved are in the order of $\chi \sim 10$ which is sufficiently large for many spin systems with local dimension $d=2$. For the 2D LGT simulations presented in this work however, we have to deal with a local dimension of $d=35$ which raises a non-trivial challenge for the PEPS ansatz. And further, this local dimension increases for going to 3D systems or a higher representation for the discretisation of the electric field.

The TTN, on the other hand, offers a more favorable computational scaling with bond-dimension: Both exact full contraction and optimization algorithms scale with $\mathcal{O}(\chi^4)$, which in turn allows
typical bond-dimensions to even exceed $\chi \geq 1000$. Moreover, a TTN is fairly straight-forward to implement and not restricted to any dimensionality of the underlying system, thus the extension to 3D systems is straight-forward. On the other hand side, TTN have showed to poorly embed the area laws in two or higher dimensions \cite{FerrisTTNAreaLaw2013}. Eventually, when increasing the system size $N=L\times L$ the TTN may fail to accurately describe the quantum wavefunction. Thus, even though the TTN is a powerful tool to tackle systems in one, two, and three dimensions, further development and improvement is needed for reaching a scalable algorithm for higher system sizes. Anyhow, since this is a variational ansatz with increasing precision for increasing bond-dimension, we can always give an estimate of the total error of our simulation results.

Our TTN algorithm implemented for finding the many-body ground state in this LGT analysis follows the prescriptions of Ref.~\cite{TTNA19}. In the numerical implementation we exploit the $U(1)$ symmetry corresponding to the conservation of total charge $Q$ using common techniques for global symmetry conservation in TNs~\cite{sv13}. We construct the tree starting from the physical indicees at the bottom by iteratively merging two local sites into one by a randomly initialized tensor coarse-grained site. In case we reach the maximum bond dimension for the coarse-grained space, we truncate the coupling symmetry sectors randomly in order to keep the bond dimension. Thereby, we randomly initialize not only the tensors themselves but the distribution of the coupling symmetry sectors within the tensors as well. In order to ensure convergence during the optimization, we dynamically increase the bond dimension locally allowing to adapt the symmetry sectors within the tree. In particular, we exploit the single tensor optimization with subspace-expansion presented in Ref.~\cite{TTNA19} which approximates a two-site update by expanding the connecting link and iteratively optimizing the two local tensors separately. Thereby, we maintain the beneficial numerical complexity of $\mathcal{O}\left(\chi^4\right)$ instead of a heavier scaling of $\mathcal{O}\left(\chi^6\right)$ for the complete two-site optimization. For the single tensor optimization we exploit the Arnoldi algorithm implemented in the ARPACK library. In this algorithm the local eigenvalue problem is solved by iteratively diagonalizing the effective Hamiltonian $H_{eff}$ for the single tensor. It delivers the lowest eigenpairs of $H_{eff}$ up to a predefined precision $\epsilon$ by requiring only knowledge of the action of the operator $H_{eff}$. In the global optimization we sweep through the TTN from the bottom to the top, performing the subspace expansion from each tensor towards its `parent' tensor (the one located directly above in the geometry of Fig.~\ref{fig:probTTN}). After one complete sweep we start over, iterating until global convergence, in terms of energy and selected observables, is achieved. As we come closer to convergence with each sweep, we as well drive the optimization precision $\epsilon$ of the Arnoldi algorithm, such that we become more and more accurate in solving the local eigenvalue problems.

The computations with TTN presented in this work ran on different HPC-cluster (the BwUniCluster and CINECA), where a single simulation of e.g. an $8x8$ system can last up to three weeks until final convergence, depending on the system parameters. Here, we point out, that we can still improve the efficiency of the code and have the potential to heavily parallelise our TTN to decrease the computational effort.

\section{Tensor network simulations for Lattice Gauge Theories}\label{app:TTN}

In this section, we describe the Tensor Network (TN) approach for LGT in more technical details. As mentioned in section \ref{sec:TNsim} we already fulfill the Gauss law by choosing the local gauge-invariant states (Sec \ref{sec:dressed}) as the logical basis in the TN simulations. 
In particular, we use a unconstrained Tree TN (TTN) to represent the many-body wavefunction~\cite{TTN14}. We adapt the TTN structure for the 2D system as shown in~\cite{gqh17, TTNvsNN}. Following the description in~\cite{TTNA19}, we additionally exploit the Abelian $U(1)$-symmetry which corresponds to the total charge $\hat Q$ for the TTN representation. In this way we keep the total charge $\hat Q$ fixed for each simulation by choosing the proper global symmetry sector.

As discussed in Sec \ref{sec:S1rep} the chosen local basis does naturally not respect the extra link symmetry arising from the division of the Hilbert space for each link into two half-links. Thus, additionally to the LGT-Hamiltonian $\hat H$ (\ref{eq:H_QED}), we include a term to penalize the states violating the link constraint during the simulation. In conclusion, we simulate the Hamiltonian
\be
\hat H_{sim} = \hat H + \nu \sum_{x,\mu} \left ( 1 - \delta_{2, \hat L_{x,\mu}} \right) ,
\label{eq:Hsim}
\ee
with $\mu \in \{\mu_x, \mu_y\}$, where the penalty term vanishes when the link symmetry is respected and increases the energy for a state breaking the symmetry. Let us mention, that this additional term translates to a nearest-neighbor interaction term in the TN simulations.

\begin{figure}[t!]
\includegraphics[width=0.24\textwidth]{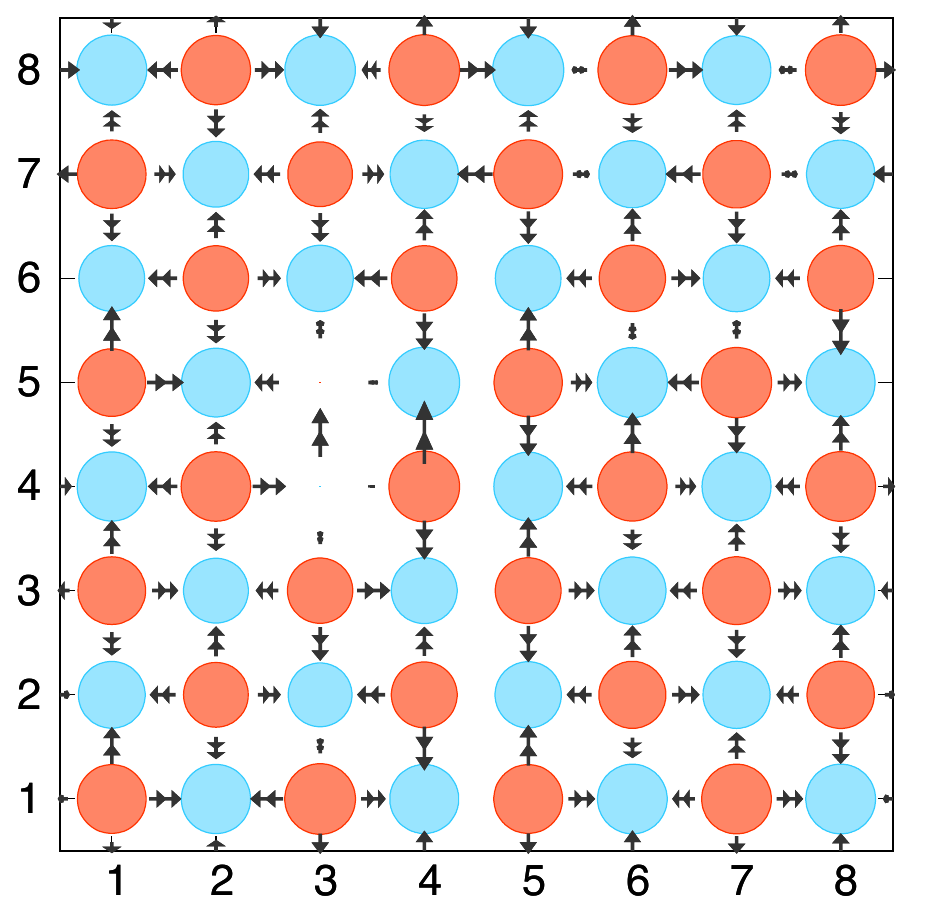}\includegraphics[width=0.24\textwidth]{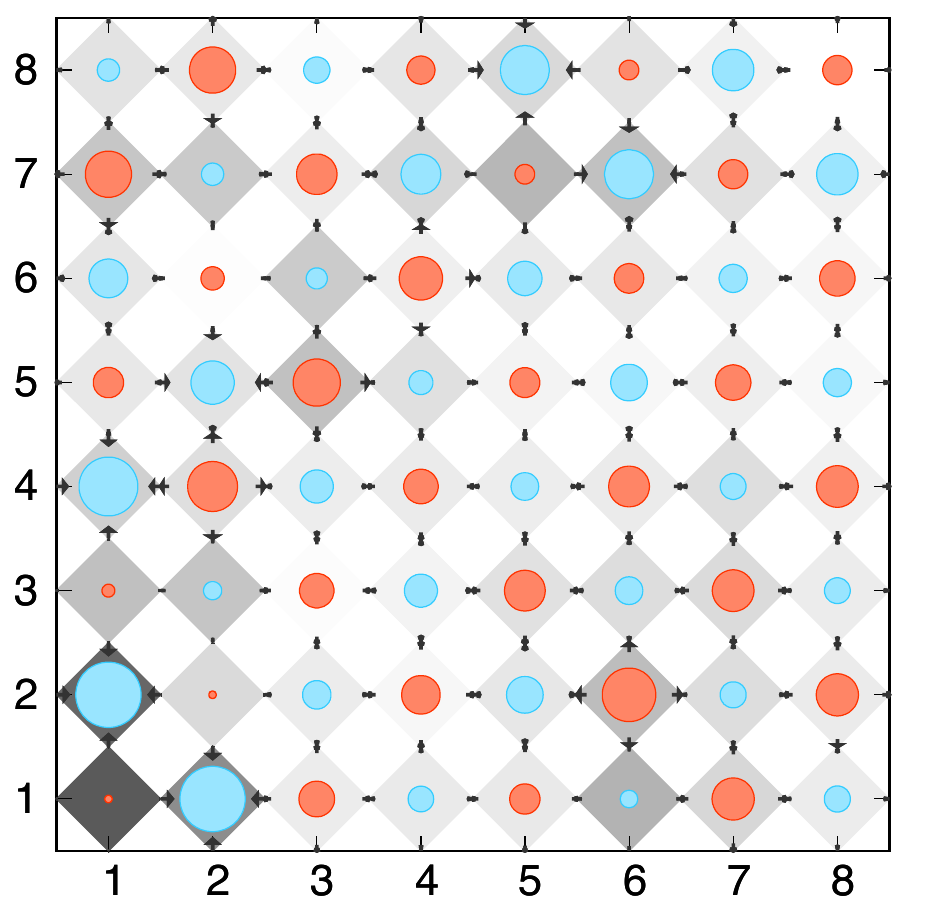}
\caption{Field plots from a TTN numerical simulations of
$8 \times 8$ systems. The left panel depicts a configuration corresponding to a local minimum in the total energy in which a simulation got stuck for a poor choice of the penalty parameter $\nu$. The right panel shows the field plot for a typical randomly initialized state. Note, that in this case the link symmetry is not respected. The gray diamonds in the background of each site signal the violation of this constraint. The darkness of the gray-color corresponds to the contribution of the penalty term in eq. (\ref{eq:Hsim}) for the site with its neighbors.
 \label{fig:FP_art_ini}
}
\end{figure}

In theory, the penalty factor $\nu$ should be chosen as large as possible, as the link symmetry is strictly enforced for $\nu\rightarrow \infty$. But choosing $\nu$ too large leads to the optimization focusing on this penalty term only and fails to optimize for the physical quantities. Depending on the physical simulation parameter $t$, $m$, $g_e$, and $g_m$ the penalty factor $\nu$ has to be chosen in a balanced way,  
such that we are able to optimize for the physical quantities as much as for the link constraint. In fact, when choosing $\nu$ too low, we end up with a result where the state does not strictly obey the link symmetry. If $\nu$ is too large,  artifacts can appear in the proposed ground state, as the penalty term can introduce local minima and thus freeze the state in the optimization. These artifacts can either be a matter-antimatter pair for the \textit{vacuum regime} or as shown on the left in Fig. \ref{fig:FP_art_ini} a matter-antimatter hole for the \textit{charge-crystal regime}. As the total charge $\hat Q$ is strictly conserved by the chosen symmetry sector during the simulation, there are only two ways to get rid of such an artifact. The optimizer has to either locally violate the link symmetry, or change the state at the neighboring sites together with the artifact - both of which would increase the energy in the simulation given a large value for $\nu$.

\begin{figure}[t!]
\includegraphics[width=0.45\textwidth]{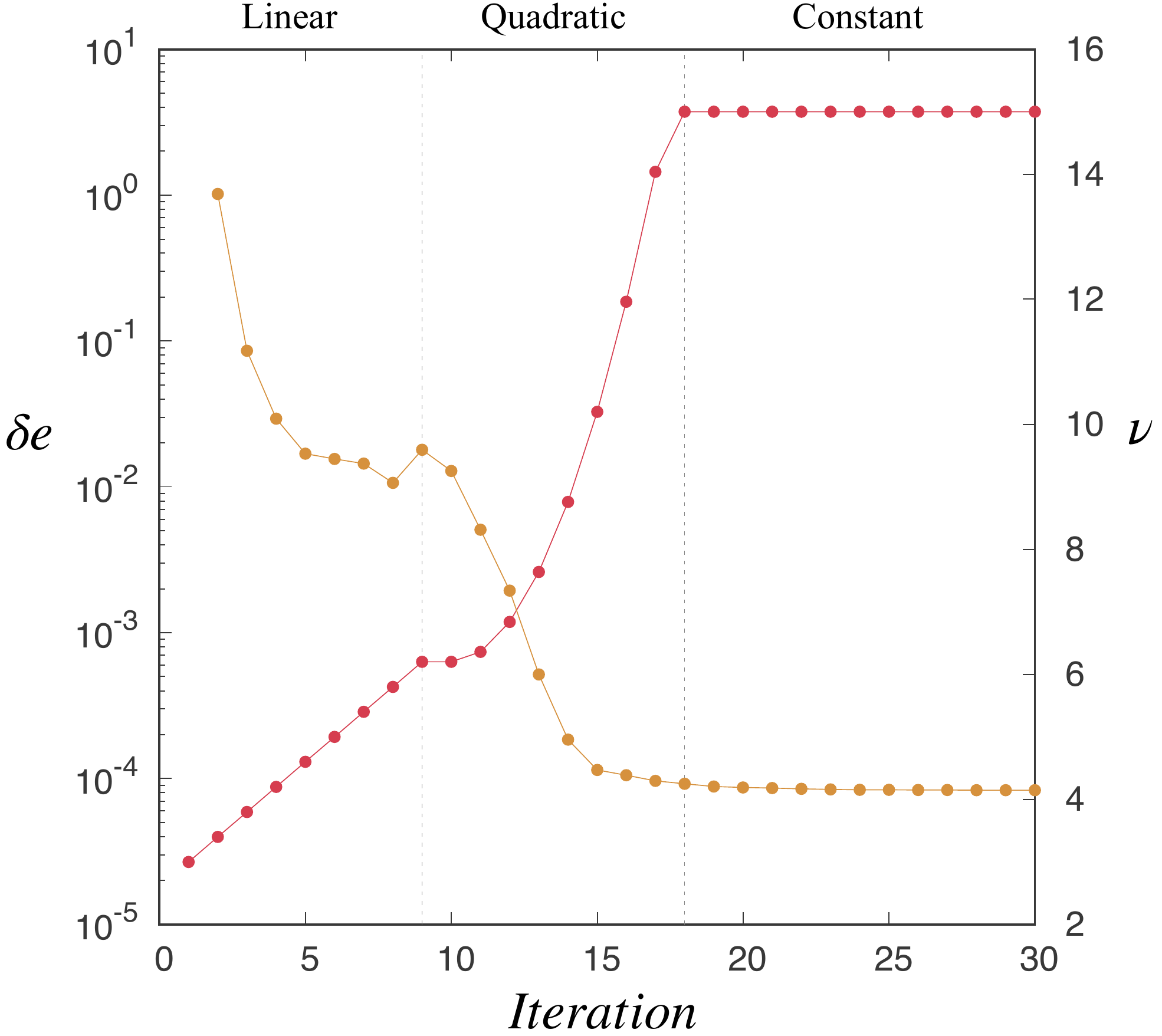}
\caption{ Penalty parameter $\nu$ (red) and energy (yellow) with respect to the number of iterations for a typical LGT simulation. The energy is plotted here as a deviation $\delta e$ to the ground state energy obtained with highest bond dimension available. We start with linearly increasing $\nu$. When the energy increases we change to a quadratic driving regime with zero gradient at the transition point. Finally we reach a predefined maximum value for $\nu$.
\label{fig:drive}
}
\end{figure}

In order to improve this approach, we exploit two different methods. First, we start with a random state, which in general violates the link symmetry. One example for this random initialization is reported on the right side of Fig. \ref{fig:FP_art_ini}. Secondly, we drive the penalty term by increasing $\nu$ after every optimization sweep. In particular, we start by linearly increasing $\nu$, until we observe an increment in the energy which signals that the penalty term becomes significant for the optimization. Consequently, we switch to a quadratic tuning of $\nu$ such that in the following few iterations we increase $\nu$ slower than in the linear regime. Finally, we as well set a maximum value for $\nu$ at which we stay for the rest of the optimization. The three different regimes of driving the penalty parameter $\nu$ are depicted in Fig.~\ref{fig:drive} showing the energy difference $\delta e$ to a higher bond dimension together with $\nu$ with respect to the iterations for an examplifying simulation.

With this driving, we optimize the random initial state in the first phase without being too strictly focused on obeying the link symmetry. This flattens the local minima arising from including the penalty in the Hamiltonian and thereby helps to converge to the global minimum. When choosing the linear tuning correctly, most physical observables are qualitatively already captured at the end of the first driving phase without strictly obeying the link symmetry. Thus the second phase enforces the link symmetry while the last phase - with a constant $\nu$ - optimizes the state for the final quantitative ground state.

Although introducing the driven penalty drastically decreases the number of simulation which are stuck in artificial configurations, this can not be completely avoided. Therefore, we simulate several samples with different random initial state. From these samples, we perform a post-selection and check whether the obtained wavefunctions are indeed physically correct ground states. We as well observe the typical convergence for TN with increasing bond dimension when we discard the results with artifacts. From the different samples and the convergence in bond dimension we can estimate the relative error in the energy which depending on the physical parameters typically lays in the range of $\sim 10^{-2}-10^{-4}$ for a $8\times 8$ system.

\section{Integral Estimators for the Correlation Length} \label{app:corlenestimator}

Here we briefly discuss a strategy to estimate correlation lengths based on integrals of the correlation functions.
The obvious advantage is that this strategy employs the whole amount of data within the correlation function itself while requiring no data
regression. Therefore it can be easily automatized and needs no careful initialization of the fit parameters for data regression,
and at the same time is a reliable, only slightly biased, estimator for correlation lengths \cite{IntegralCorrelationLengthWhiteDMRG}.

While in the main text we applied an analogous estimator to the full correlation function, in this section we perform an estimator analysis
on the connected component
$\mathcal{C}^{0}_{x,x'} = \langle {\hat{\mathcal{O}}}_{x} {\hat{\mathcal{O}}}_{x'} \rangle - \langle {\hat{\mathcal{O}}}_{x} \rangle \langle {\hat{\mathcal{O}}}_{x'} \rangle$
of the correlation function,
which we spatially average to $\bar{C}^{0}_v = L^{-2} \sum_{x} C^{0}_{x,x+v}$.
In the absence of strong quantum correlations, lattice systems at low temperatures typically exhibit 
$\bar{C}^{0}(v) \simeq \alpha_0 \exp(-|v|/\xi)$
exponentially decaying in the relative coordinate modulus $|v| = \sqrt{v_x^2 + v_y^2}$, where $\xi$ is the actual correlation length
and $\alpha_0$ a (not intesting) prefactor. Here we construct an integral estimator $\xi_{\text{est}}$ for (connected) correlation lengths, and show
that on the exponentially-decaying class it returns $\xi$ to an acceptable precision.

In deriving these expressions,
we assume that the system is much larger than the correlation length $L \gg \xi$ to avoid observing finite-size or boundary effects
(we effectively approximate the lattice to $\mathbb{Z}^2$).
For a 2D square lattice, we consider the following estimator
\begin{equation} \label{eq:2dintegral}
 \xi^2_{\text{est}} = \frac{\sum_{v_x,v_y \in \mathbb{Z}} |v|^2 \bar{C}^{0}(v)}{6 \sum_{v_x,v_y \in \mathbb{Z}} \bar{C}^{0}(v)},
\end{equation}
which, apart from the $1/6$ prefactor, is the (euclidean) variance of
$P(v) = \bar{C}^{0}(v) (\sum_{v'} \bar{C}^{0}(v') )^{-1}$, the correlation function normalized to a probability distribution over $\mathbb{Z}^2$
(assuming $\bar{C}^{0}(v)$ is symmetric $\bar{C}^{0}(v) = \bar{C}^{0}(-v)$).

In the limit of correlation lengths large compared to the lattice spacing, $\xi \gg 1$, the discrete sums in Eq.~\eqref{eq:2destimator} converge
to Riemann integrals
\begin{equation} \label{eq:2destimator}
 \xi^2_{\text{est}} = \frac{\int_{\mathbb{R}^2} |v|^2 \bar{C}^{0}(v) {d^2}v}{6 \int_{\mathbb{R}^2} \bar{C}^{0}(v) {d^2}v},
\end{equation}
yielding an unbiased estimator $\xi^2_{\text{est}} = \xi^2$ for the family of correlation functions $\bar{C}^{0}(v) \simeq \alpha_0 \exp(|v|/\xi)$.
For correlation lengths comparable in magnitude with the lattice spacing, the finite sum in Eq.~\eqref{eq:2destimator} can produce a bias
$B(\xi) = \xi^2 - \xi^2_{\text{est}} \geq 0$ in the estimator. Unfortunately, $B(\xi)$ is not an analytic function and thus can not be removed
altogether. However, we numerically verified that $B(\xi)$ is upper bounded by $1/17$, for any $\xi \in \mathbb{R}$, which makes
Eq.~\eqref{eq:2destimator} a satisfactory estimator for the purposes of identifying phases and transitions.

\section{Boundary Hamiltonian and typical boundary conditions}\label{app:boundary}

In this section, we discuss strategies to realize a specific set of (open) boundary conditions for problems of equilibrium electrodynamics.
These strategies present an extension to the simulations realized in this work, who assume the boundary conditions to be either
free (for finite charge density) or periodic (for zero charge density).

{\it Von Neumann boundary conditions} $-$ In this simple scenario, the outgoing electric flux at each boundary site is fixed and defined by the user.
To realize it, start the TTN algorithm from a product state that has the desired configuration of electric fluxes
at the open boundary, and then simply carry out the optimization algorithm (without a boundary Hamiltonian, i.e.~$H_b = 0$). The algorithm has no means of changing
the electric fluxes at the boundaries, and will converge to the bulk ground state given that specific boundary flux configuration.

{\it Dirichelet boundary conditions} $-$ To model the scenario where the boundaries are a perfect conductor, we actually assume the boundaries to be superconductive, and expel magnetic fields by displaying huge magnetic couplings at the boundary. This requires the usage of a magnetic boundary Hamiltonian
\begin{equation}
\begin{aligned}
 H_b &= J_b \left( \sum_{j=1}^{L-1} \left( \vphantom{\hat U^{\dag}_{(1,j),-\mu_{x}}} \right. \right.
\hat U^{\dag}_{(1,j),-\mu_{x}}  \hat U_{(1,j),\mu_{y}} \hat U_{(1,j+1),-\mu_{x}} \\
&+ \hat U^{\dag}_{(j,L),\mu_{y}}  \hat U_{(j,L),\mu_{x}} \hat U_{(j+1,L),\mu_{y}} \\
&+ \hat U^{\dag}_{(L,j+1),\mu_{x}}  \hat U^{\dag}_{(L,j),\mu_{y}} \hat U_{(L,j),\mu_{x}} \\
&+ \left. \hat U^{\dag}_{(j+1,1),-\mu_{y}}  \hat U^{\dag}_{(j,1),\mu_{x}} \hat U_{(j,1),-\mu_{y}} \right) \\
&+ U^{\dag}_{(1,L),-\mu_{x}}  \hat U_{(1,L),\mu_{y}} + U^{\dag}_{(L,L),\mu_{y}}  \hat U_{(L,L),\mu_{x}} \\
&+ U_{(L,1),\mu_{x}}  \hat U^{\dag}_{(L,1),-\mu_{y}} + U^{\dag}_{(1,1),-\mu_{y}}  \hat U_{(1,1),-\mu_{x}} 
\\& \left. + \text{H.c.} \vphantom{\sum} \right),
\end{aligned}
\end{equation}
which contains both edge terms (top rows) and corner terms (bottom rows). To address the problem of electrodynamics the ground state algorithm
is carried out while setting $J_b \gg \max\{|t|, |m|, g^2_e, g^2_m\}$, ensuring that the magnetic fields will approach a constant value (equal to zero) at
the boundary, once converged.

\section{Perturbation theory}\label{app:perturbation}
Here we describe the corrections to the ground state 
in both the two regimes outlined in Section~\ref{sec:zero_charge}.
Let us start by considering particles fluctuations due to the presence of a small tunneling $|t|$.
The system has periodic boundary conditions.

\begin{figure}[t!]
\includegraphics[width=0.49\textwidth]{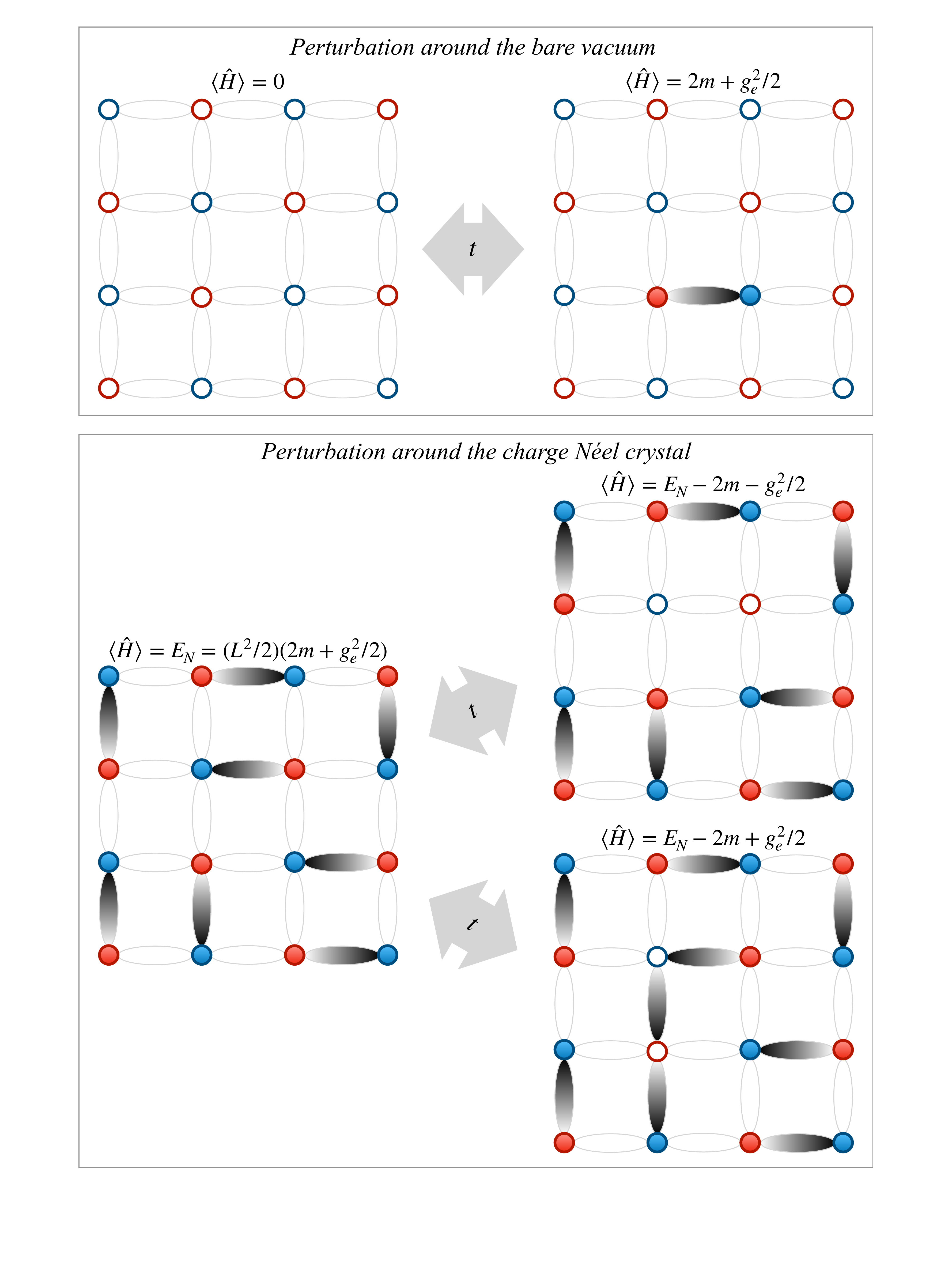}
\caption{ \label{fig:perturbation}
Example of excited states coupled to the vacuum (top) 
or to the fully dimerised state (bottom)
at the lowest order in perturbation theory in the tunneling coupling $t$,
as described in Appendix~\ref{app:perturbation}.
}
\end{figure}

\paragraph{Perturbation around the vacuum state.---}
For $m\gg |t|$, the vacuum state 
(with zero energy) is corrected by 
strictly local particle-antiparticle fluctuations. 
The first nontrivial contribution 
comes from a local dimer excitation as depicted in Fig.~\ref{fig:perturbation}, 
whose average energy is $2m+g^{2}_{e}/2$. 
The truncated Hamiltonian reads (a part from the sign of the tunneling coupling,
which however does not affect the results)
\be
H_{v} =
\left[
\begin{array}{c|ccc}
0  & t & \cdots & t \\
\hline
t & 2m+g^{2}_{e}/2 & & \\
\vdots & & \ddots & \\
t & & & 2m+g^{2}_{e}/2 \\
\end{array}
\right],
\ee
which is $(1+2L^2) \times (1+2L^2)$ matrix.
The correction to the vacuum energy is therefore
\be
E_{v} =
\left[
\frac{g^{2}_{e}}{4}+m-\sqrt{\left(\frac{g^{2}_{e}}{4}+m \right)^{2}+2L^2 t^2}\,
\right].
\ee

\paragraph{Perturbation around the dimer state.---}
Small-order tunneling perturbations on top of the fully dimerised states
are not sufficient to remove their degeneracy. The ground-state energy sector
remains degenerate up to the fourth-order in perturbation theory.
Here we focus on the smallest order energy corrections 
for one specific dimerised configuration, and consider the possible excitations
as depicted in Fig.~\ref{fig:perturbation}. We now have two different
excitation sectors, depending where we remove a particle/antiparticle pairs:
when the pairs is annihilated on top of a dimer, the energy cost is $2m + g^{2}_{e}/2$;
otherwise, when we remove a pairs in between two dimers, we have to spend
$2m - g^{2}_{e}/2$. The number of possible configurations of the first type
coincides with the number of dimers, i.e. $L^2/2$; in the other case, we 
have $3L^2/2$ different possibilities. The full truncated Hamiltonian
is still a $(1+2L^2) \times (1+2L^2)$ matrix which now reads
(a part from the overall extensive constant $E_{N}\equiv(2m + g^{2}_{e}/2)L^2/2$)

\begin{widetext}
\be
H_{d} =
\left[
\begin{array}{c|ccc|ccc}
0  & t & \cdots & t & t & \cdots & t \\
\hline
t & -2m-g^{2}_{e}/2 & & & & & \\
\vdots & & \ddots & \\
t & & & -2m-g^{2}_{e}/2 \\
\hline
t & & & & -2m+g^{2}_{e}/2 & & \\
\vdots & & & & \ddots & \\
t & & & & & & -2m+g^{2}_{e}/2
\end{array}
\right],
\ee
\end{widetext}

where also in this case the sign of $t$ does not affect the results. 
The correction to the vacuum energy can be evaluated as well
by solving $\det(H_{d}-\varepsilon) = 0$; indeed, due to the structure of the 
matrix, and thanks to the properties of the determinant,
we found 
\be
E_{d} = \frac{L^2}{2} \left( \frac{g^{2}_{e}}{2} + 2m  \right) + \varepsilon_{-},
\ee
where $\varepsilon_{-}$ is the negative solution of
\be\label{eq:dimer_energy_corr}
\varepsilon
\left[ \frac{g^{4}_{e}}{4} - (2m  +\varepsilon)^2 \right]
+ 2L^2 t^2 
\left(
 \frac{g^{2}_{e}}{4} +2m + \varepsilon
\right)
= 0 .
\ee

Now it is clear that, when $m$ is approaching the value $-g^{2}_{e}/4$, 
the biggest corrections, at the lower order in $t$, solely come from the sector quasi-degenerate with
the classical dimerised configuration.
The finite-size scaling depends whether the mass is approaching from above (i.e. from the vacuum)
or form below (i.e. from the dimerised configuration): in the first case $2L^2$ states contributes to the 
energy corrections; in the second case, if $g_{e} > 0$, only $L^2/2$ states get involved. An energy 
gap ${|E_{v}-E_{d}| \sim L t /\sqrt{2}}$ opens. Notice that, in the pathological situation
where $g_{e} = 0$ as well, there is no gap opening at the second order in $t$ and therefore
a sharper transition is expected. 

\begin{figure}[t!]
\includegraphics[width=0.49\textwidth]{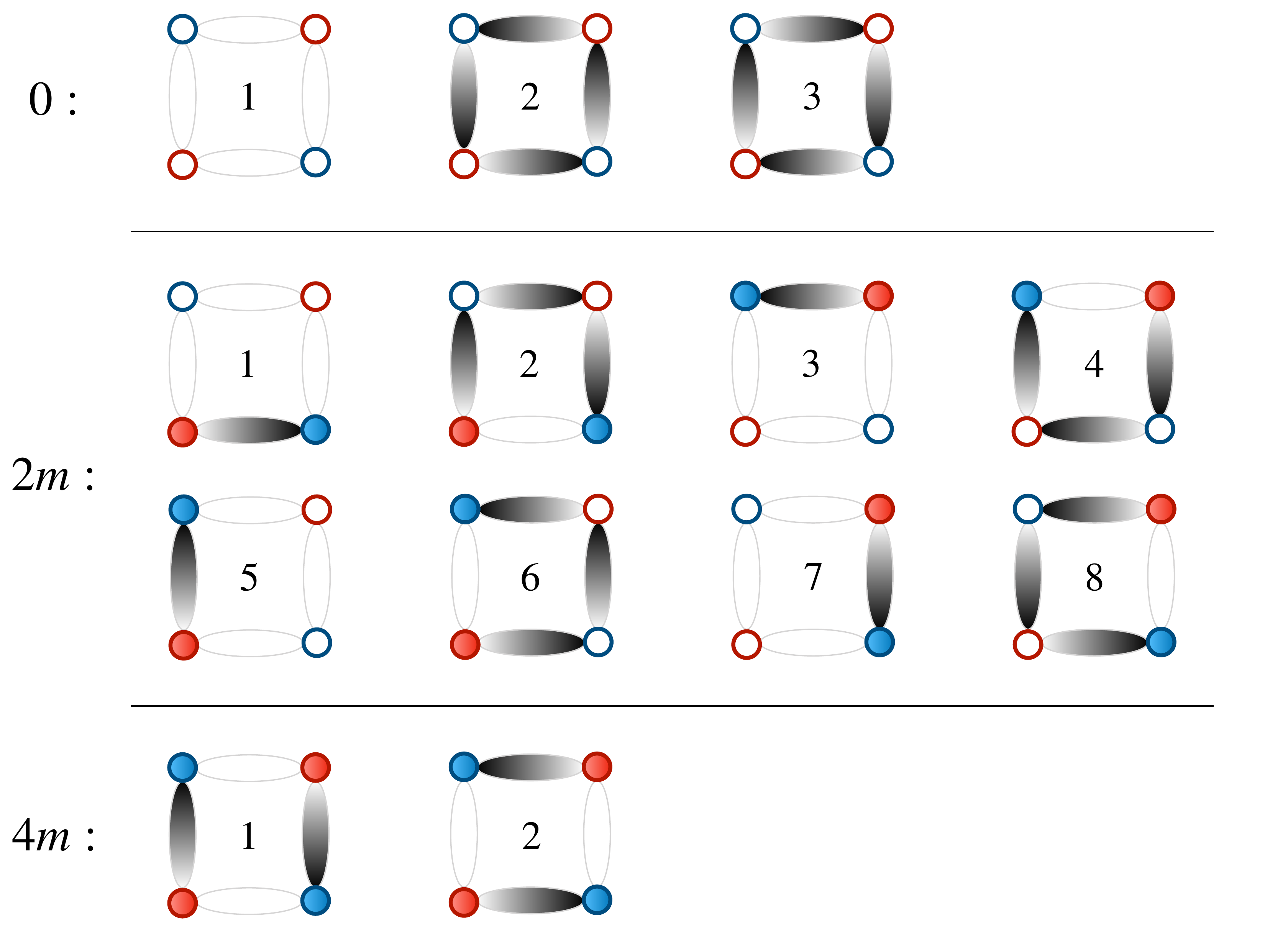}
\caption{ \label{fig:2x2_basis}
Graphic representation of the basis vectors, in each mass sector, used to build up the 
$2 \times 2$ LGT Hamiltonian in the $S=1$ representation, 
as outlined in Appendix~\ref{app:2x2}.
}
\end{figure}

Let us mention that, the correction to the ground-state energy coincides, as it should, with the
second-order degenerate perturbation theory. In practice, if $\hat {\mathcal Q}$ is the projector into the
classical charge-crystal sector, and $\hat {\mathcal P} = 1-\hat {\mathcal Q}$ the projector into the complementary sector,
then we may split the eigenvectors in two contributions:
$|E_{k}\rangle = |\phi_{k}\rangle + |\varphi_{k}\rangle$, 
where $ |\phi_{k}\rangle \equiv \hat {\mathcal Q} |E_{k}\rangle$
and
$ |\varphi_{k}\rangle \equiv  \hat {\mathcal P} |E_{k}\rangle$.
The eigenvalue equation 
$(\hat H_{0}-t\hat V) |E_{k}\rangle = E_{k} |E_{k}\rangle$
therefore splits in two coupled equations 
\bea
-t  \hat {\mathcal P} \hat V |\phi_k\rangle & = & (E_{k} - \hat H_{0}+t \hat {\mathcal P} \hat V \hat {\mathcal P}) |\varphi_{k}\rangle \\
-t \hat {\mathcal Q} \hat V |\varphi_k\rangle & = & (E_{k} - E_{N}) |\phi_{k} \rangle
\eea
where we used the fact that, in our case $\hat {\mathcal Q} \hat V \hat {\mathcal Q} = 0$.
Corrections within the degenerate sub-sector are thus given by recursively solving the following equation
\be
(E_{k} - E_{N}) |\phi_{k} \rangle = 
\hat {\mathcal Q} \hat V \hat {\mathcal P}
\frac{t^2}{E_{k}-\hat H_0+ t \hat {\mathcal P}\hat V \hat {\mathcal P}}
\hat {\mathcal P} \hat V 
 |\phi_{k} \rangle.
\ee
At the second order in the tunneling, the dimerised sub-sector degeneracy is not lifted,
and the energy changes according to Eq. (\ref{eq:dimer_energy_corr}).
Let us stress that, when deep into the charge-crystal regime, these are the dominant corrections.
However, close to the classical transition, the creation/annihilation of particle-antiparticle
is energetically favorable, and non-trivial corrections to the degeneracy of the ground-state energy sector
are induced by fourth-order tunneling transitions: two different classical dimerised states are 
coupled whenever they share at least one ``resonating'' {\it plaquette}, 
which consists in two neighboring horizontal/vertical dimers 
(see the $4m$ mass sector in Fig.~\ref{fig:2x2_basis}). This effect 
partially removes the ground-state degeneracy, making energetically favorable
a specific superposition of different dimer states. Incidentally, let us mention that,
in the thermodynamic limit, there exist classical dimer configurations,
e.g. the state where dimers are all vertically (horizontally) aligned  
with all local electric fluxes pointing in the same direction, 
which are not resonating with any other fully dimerised state 
at any order in perturbation theory.
 
\section{Exact results of the $2 \times 2$ system}\label{app:2x2}

In the zero-charge density sector, the single plaquette system, i.e. $2 \times 2$,
admits only $13$ gauge-invariant diagonal configuration,
in the Spin-$1$ compact representation of the electric field.
The full Hamiltonian can be easily constructed
by considering each mass sector $\{0,2m, 4m\}$ independently,
and it acquires the following block structure,
\be \label{eq:H2x2}
H_{2\times 2} = 
\left[
\begin{array}{c|c|c}
D_{0} & T_{02} & \emptyset \\
\hline
T_{20}& D_{2} & T_{24} \\
\hline
\emptyset & T_{42}& D_{4}
\end{array}
\right] ,
\ee
where $D_{j} = D^{\dag}_{j}$, 
$T_{20}= T^{\dag}_{02}$,  $T_{42}= T^{\dag}_{24}$,
and all matrix entries are reals. To construct each block,
we used the gauge-invariant eigenstates of the electric field $\hat E_{x,\mu}$
and particle number $\hat n_{x}$, as listed in Fig.~\ref{fig:2x2_basis}.

\begin{figure}[t!]
\includegraphics[width=0.49\textwidth]{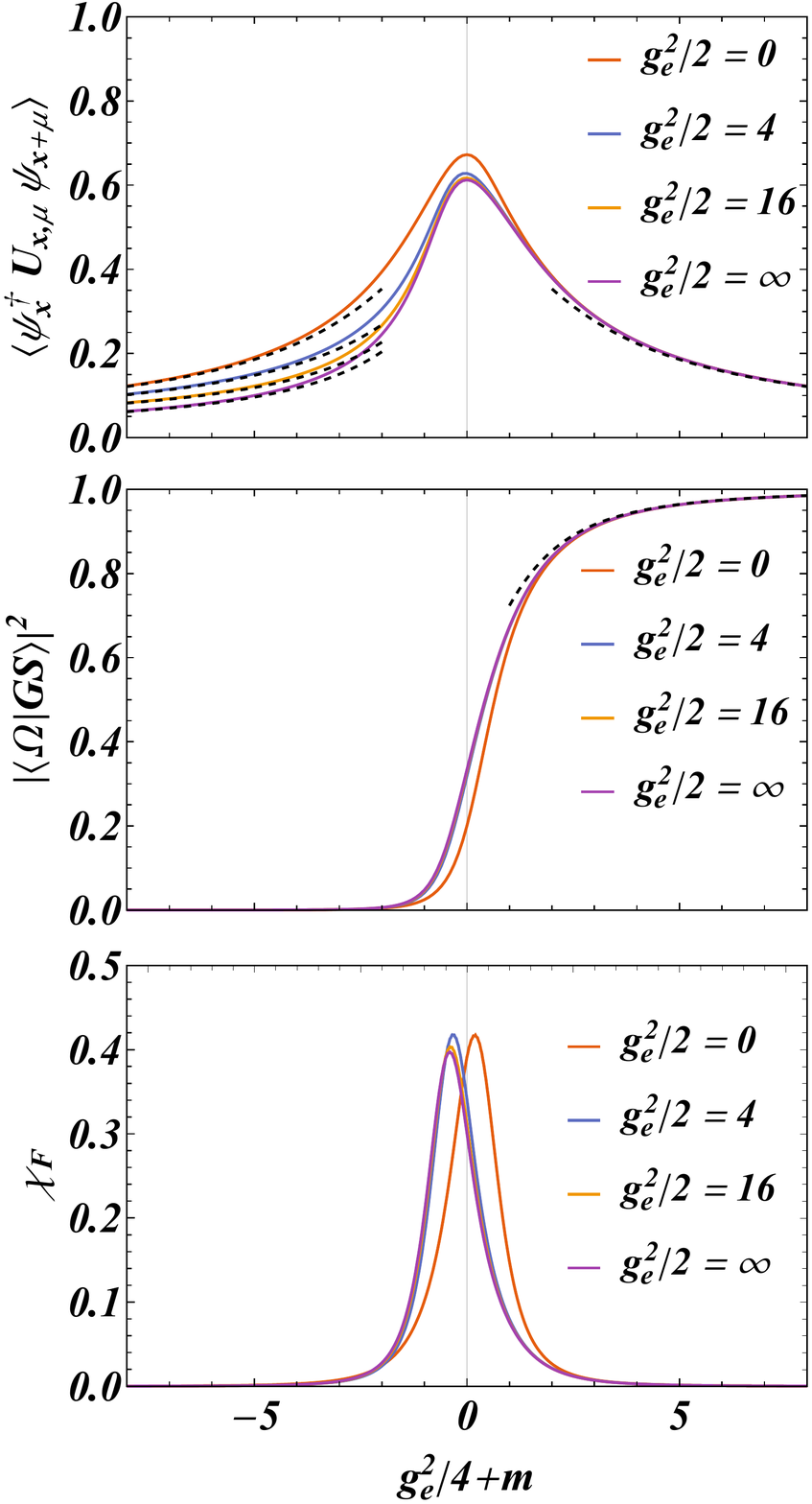}
\caption{\label{fig:2x2_observables}
({\bf top}) Expectation value of the tunneling Hamiltonian
as a function of the distance from the classical transition for a $2 \times 2 $ system. 
({\bf centre})The module square of the overlap between the exact $2\times 2$ ground-state and the vacuum
sate when varying the coupling across the classical transition point. 
Dashed lines are the perturbative predictions. 
({\bf bottom}) The fidelity susceptibility of the ground state
as defined in the main text.
}
\end{figure}

The diagonal blocks read
\be
D_{0}  = 
\begin{pmatrix}
0 & -g^{2}_{m}/2 & -g^{2}_{m}/2 \\
-g^{2}_{m}/2 & 2 g^{2}_{e} & 0 \\
-g^{2}_{m}/2 & 0 & 2 g^{2}_{e}
\end{pmatrix} ,
\ee

\be
D_{2}  = 
\mathbb{I}_{4}
\otimes
\begin{pmatrix}
2m + g^{2}_{e}/2 & -g^{2}_{m}/2  \\
-g^{2}_{m}/2 & 2m + 3g^{2}_{e}/2 
\end{pmatrix} ,
\ee

\be
D_{4}  = 
\begin{pmatrix}
4m + g^{2}_{e} & -g^{2}_{m}/2  \\
-g^{2}_{m}/2 & 4m + g^{2}_{e}
\end{pmatrix} ,
\ee
where $\mathbb{I}_{4}$ is a $4 \times 4$ identity matrix.
The out-diagonal blocks are responsible for creation/annihilation of
particle/antiparticle pairs and are given by
\be
T_{02}  =
\begin{pmatrix}
-t & 0 & -t & 0 & t & 0 & -t & 0 \\
0 & 0 & 0 & 0 & 0 & t & 0 & -t \\
0 & -t & 0 & -t & 0 & 0 & 0 & 0 
\end{pmatrix} , 
\ee
\be
T_{42}  = 
\begin{pmatrix}
0 & -t & 0 & -t & -t & 0 & -t & 0 \\
-t & 0 & -t & 0 & 0 & -t & 0 & -t 
\end{pmatrix} .
\ee

The exact diagonalisation of the Hamiltonian $\hat H_{2 \times 2}$
allows us to explore the behavior of the ground state in the vicinity of the
transition $m \simeq g^{2}_{e}/4$. As expected from the enhancement of quantum fluctuations,
the gauge-invariant hopping term gets picked at the transition (Fig.~\ref{fig:2x2_observables} top panel).
The overlap of the ground state with the bare vacuum as function of $m$ for different values of the electric coupling 
is analyzed as well  (central panel in Fig.~\ref{fig:2x2_observables}).
Exact curves are compared with first-order perturbative results. 

\begin{figure}[t]
\includegraphics[width=0.4\textwidth]{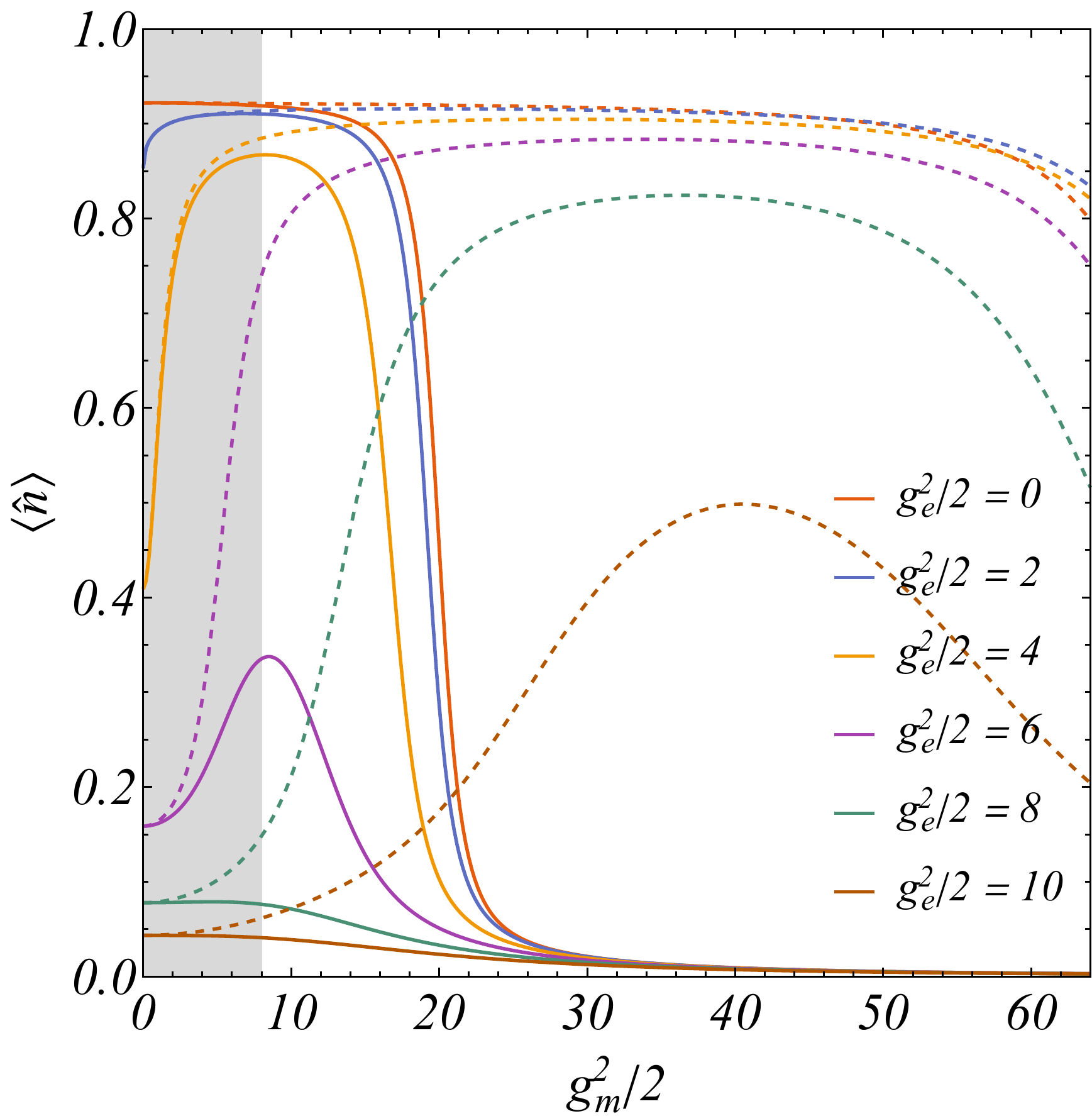}
\caption{ \label{fig:2x2_magnetic}
Behavior of the particle density vs the magnetic coupling 
in the $2 \times 2$ system for $m=-2$ and different electric couplings $g^{2}_{e}/2$.
The shaded gray area represents the region explored in Fig.~\ref{fig:magnetic_vac}.
Full lines are the $S=1$ results; dashed lines are the $S=2$ results.}
\end{figure}

In order to explore with more care the transition region, we 
look at the fidelity susceptibility of the ground state~\cite{gu10,damski13,brau18,rv18},
$
\chi_F(m) \equiv \langle \partial_{m} GS(m) | \partial_{m} GS(m) \rangle 
- | \langle GS(m) | \partial_{m} GS(m) \rangle |^2 ,
$
which gives the leading contribution to the ground state fidelity
$
|\langle  GS(m) | GS(m+\delta) \rangle|  = 1 - \delta^2 \chi_{F}(m)/2 + o(\delta^2),
$
since the linear contribution in $\delta$ vanishes 
due to the normalization condition $\langle GS(m)| GS(m)\rangle = 1$.
This quantity is the perfect indicator of a changing in the geometrical properties of the ground state
when varying the couplings. Moreover, from perturbation theory, it can be easily shown that
$
\chi_{F}(m) \leq 
[\langle GS(m)| (\sum_x \hat n_{x})^2 |GS(m)\rangle-\langle GS(m)| \sum_x \hat n_{x} | GS(m)\rangle^2]/\Delta^{2}
$, 
where $\Delta$ is the energy gap between the 
ground state and the lower excitations. In practice, 
the fidelity susceptibility of the ground state is bounded from above by the 
number of particle fluctuations (which is an extensive quantity)
divided by the gap. Whenever $\chi_{F}(m)$ shows a super-extensive behavior, 
the ground state of the system should be gapless.
From the numerical data we have confirmation that $\chi_{F}(m)$ 
is enhanced in the vicinity of the transition between the two 
regions, as depicted in the bottom panel of Fig.~\ref{fig:2x2_observables}.

\begin{figure}[t]
\includegraphics[width=0.45\textwidth]{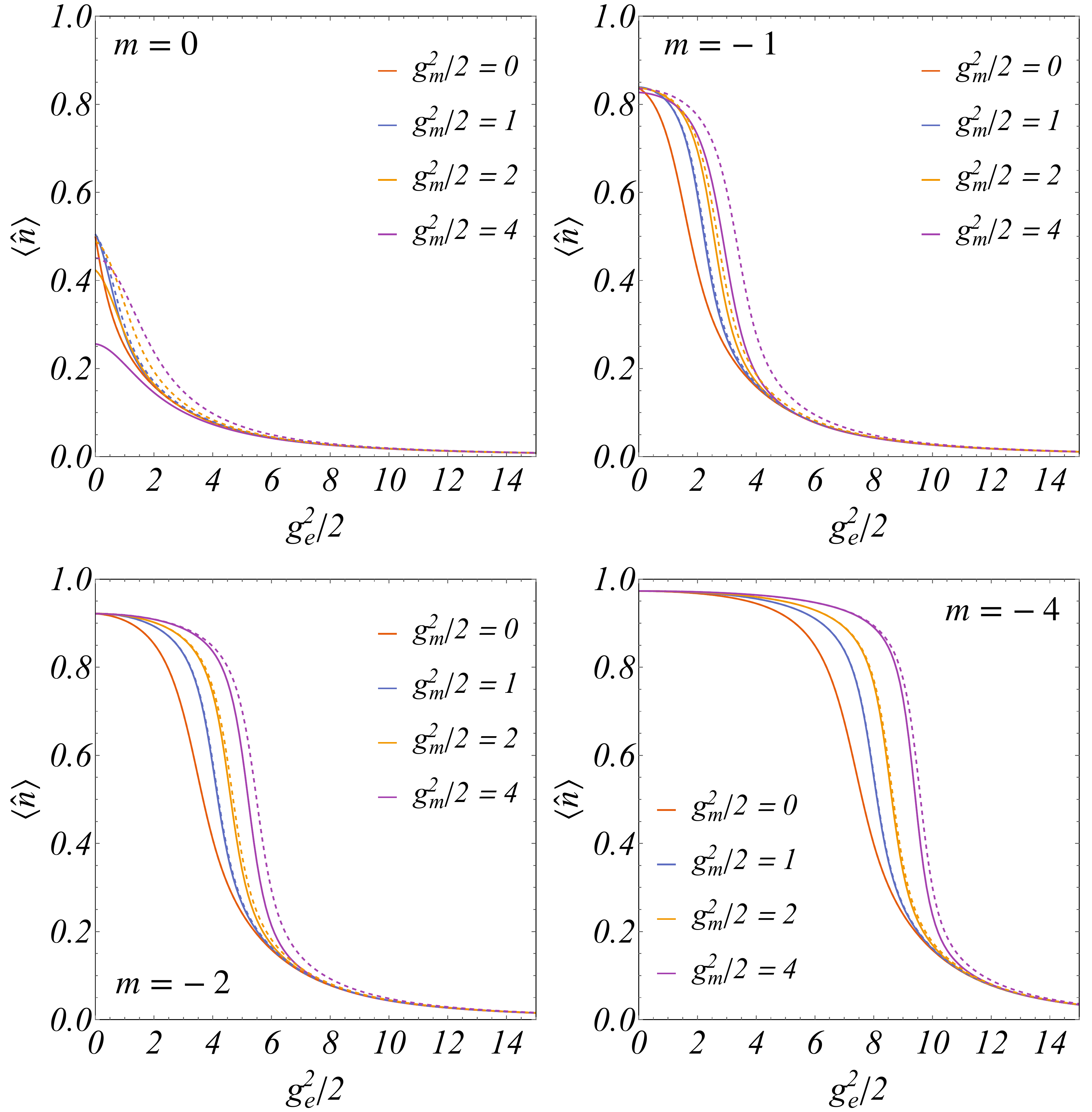}
\caption{ \label{fig:2x2_S1vsS2}
Particle density as a function of the electric coupling $g_e$
for different values of the magnetic coupling $g_m$. 
The four panels represent different bare masses.
Full lines are the $S=1$ results; dashed lines are the $S=2$ results.}
\end{figure}

In Fig.~\ref{fig:2x2_magnetic} we reproduce the 
 behavior of the matter density as function of the magnetic coupling,
 for different values of the electric field couplings. As explained in the main text,
 and confirmed by these exact results in the $2\times 2$ {\it plaquette},
 the local density gets enhanced by applying a small magnetic coupling; 
 however, when $g^{2}_{m} \simeq g^{2}_{e}$, 
 the particle density starts decreasing and eventually vanishing for $g^{2}_{m}\gg  g^{2}_{e}$. 
 Let us stress that this phenomenon is strictly due to the finite compact representation of the gauge field.

Indeed, when gauge-field fluctuations are very strong, 
 we may expect deviation in the observables 
 due to the finite Spin representation of the electric field;
 in order to have an estimate of the finite-$S$ representation accuracy,
 we further analyze the $2\times 2$ plaquette system in the $S=2$ compact representation,
 namely when the electric field (in unit of flux) can get the values $\{-2,-1,0,1,2\}$.
 The full Hamiltonian still preserves the block structure in Eq. (\ref{eq:H2x2}),
 where now each mass sector acquires further gauge-invariant states, 
 for a total of: $5$ states in the $0$-mass sector;
 $16$ states in the $2m$-mass sector;
 $4$ states in the $4m$-mass sector.
 
 As a matter of fact, when $S=2$, the phenomenon of density suppression
 depicted in Fig.~\ref{fig:2x2_magnetic} occurs for much larger values of the
 magnetic couplings, thus disappearing in the limit $S\to\infty$.
 
 In Fig.~\ref{fig:2x2_S1vsS2} we compare the matter density
 for the two compact representations $S=1,2$ and
 different values of the couplings.
 As expected, for $g_e^2 \gg g_m^2$ the two representations are equivalent;
moreover, if $-m \gg1$ (i.e. very negative) the diagonal configurations are more energetically favorable
and even for small electric coupling, and a finite value of $g_m^2$, 
the truncation of the gauge-field representation  
does not affect too much the results ($S=1$ and $S=2$ are almost identical indeed);
of course, for $m \geqslant 0$ this is not the case and we need $g_{e}^2 \gg g_{m}^2$:
notice that, in the actual QED this condition is satisfied 
as far as the electric coupling is sufficiently large, since 
$g_e^2 \sim g_m^{-2} \sim g^2$.

\newpage



%

\end{document}